\documentclass[conference]{ieeeconf}

\IEEEoverridecommandlockouts  
\usepackage[utf8]{inputenc} 
\usepackage[T1]{fontenc}    
\usepackage{hyperref}       
\usepackage{url}            
\usepackage{booktabs}   
\usepackage{amsmath}
\usepackage{amsfonts}
\usepackage{xcolor}
\usepackage{soul}
\usepackage{nicefrac}       
\usepackage{microtype}      
\usepackage{lipsum}
\usepackage{fancyhdr}       
\usepackage{graphicx}       
\graphicspath{{media/}}     
\usepackage{caption}
\usepackage{subcaption}
\usepackage{stmaryrd}

\usepackage{algorithm} 
\usepackage{algpseudocode} 

\usepackage[utf8]{inputenc} 
\usepackage[T1]{fontenc}    
\usepackage{hyperref}       
\usepackage{url}            
\usepackage{booktabs}   
\usepackage{amsmath}
\usepackage{amsthm}
\usepackage{amsfonts}
\usepackage{graphicx}
\usepackage{xcolor}
\usepackage{epstopdf}
\usepackage{amssymb}
\usepackage{siunitx}
\usepackage{soul}
\usepackage{nicefrac}       
\usepackage{microtype}      
\usepackage{lipsum}
\usepackage{fancyhdr}  
\usepackage[ragged]{sidecap}    

\usepackage{graphicx}       
\graphicspath{{media/}}     
\usepackage{caption}
\usepackage{subcaption}
\usepackage{stmaryrd}

\usepackage{algorithm} 
\usepackage{algpseudocode}

\newtheoremstyle{thmv2}
{}
{.1em}
{\itshape}
{}
{\bfseries}
{.}
{.1em}
{}
\theoremstyle{thmv2}

\newtheorem{assumption0}{Assumption}
\newtheorem{remark}{Remark}
\usepackage{mathtools}
\usepackage{tikz}
\usetikzlibrary{shapes,arrows}
\usetikzlibrary{intersections}
\usepackage{cancel}
\usepackage[mathscr]{eucal}
\usepackage{algpseudocode}

\newcommand{\until}{\mathbin{\sf U}}

\newcommand{\Neig}{\mathbf{N}}

\newcommand{\spaceX}{\mathbb{X}} 
\newcommand{\spaceS}{\mathbb{S}}

\newcommand{\x}[1]{x_{#1}} 
 
\newcommand{\mbP}{\mathbf{P}}
\newcommand{\execution}{\boldsymbol{s}}

\newcommand{\w}[1]{w_{#1}} 

\newcommand{\mdpM}{\mathbf{M}}
\newcommand{\Tr}{{\mathbb T}}
\newcommand{\vs}{s}
\newcommand{\AP}{\mathrm{AP}}
\newcommand{\letter}{l}
\newcommand{\word}{\boldsymbol{l}}

\newcommand{\DFA}{\mathcal A}
\newcommand{\DFAs}{{\DFA_{\phi}}}
\newcommand{\setnodes}{\mathrm{N}}
\newcommand{\spaceA}{\mathbb{A}} 
\newcommand{\vac}{a}

\newcommand{\subphi}{\alpha}

\newcommand{\valuemapping}{v}

\DeclareMathOperator*{\argmax}{arg\,max}

\newcommand{\tensorP}{\mathscr{P}}
\newcommand{\tensorV}{\mathscr{V}}
\newcommand{\qmapping}{\mathcal{L}_Q}

\newcommand{\optree}{\mathcal{T}}
\newcommand{\expectation}{\mathbb{E}}
\newcommand{\indicator}{\boldsymbol{1}}

\newcommand{\outerproduct}{\otimes}
\newcommand{\op}{\mathbf{T}}

\newcommand{\new}[1]{{\color{cyan}#1}}

\newtheorem{definition}{Definition}
\newtheorem{example}{Example}

\newtheorem{lemma}{Lemma}
\newtheorem{theorem}{Theorem}

\newtheorem{proposition}{Proposition}

\begin{document}
\title{Unraveling tensor structures in correct-by-design controller synthesis}
\author{Ruohan Wang$^{1}$, Zhiyong Sun$^{1,2}$ and Sofie Haesaert$^{1}$ \thanks{This work is supported by the Dutch NWO Veni project CODEC
under grant number 18244 and the European project SymAware
under the grant number 101070802. $^{1}$Department of Electrical Engineering, Control
Systems Group, Eindhoven University of Technology, The Netherlands. $^{2}$Department of Mechanics and Engineering Science $\&$ State Key Laboratory for Turbulence and Complex Systems, Peking University, Beijing, China. Emails: $\{$r.wang2,s.haesaert$\}$@tue.nl; zhiyong.sun@pku.edu.cn.}
}

\maketitle
\begin{abstract}

Formal safety guarantees on the synthesis of controllers for stochastic systems can be obtained using correct-by-design approaches. These approaches often use abstractions as finite-state Markov Decision Processes (MDPs). As the state space of these MDPs grows, the curse of dimensionality makes the computational and memory cost of the probabilistic guarantees, quantified with dynamic programming, scale exponentially. In this work, we leverage decoupled dynamics and unravel, via dynamic programming operations, a tree structure in the Canonical Polyadic Decomposition (CPD) of the value functions.
For discrete-time stochastic systems with syntactically co-safe linear temporal logic (scLTL) specifications, we provide provable probabilistic safety guarantees and significantly alleviate the computational burden. We provide numerical studies and validations of the theoretical results on several typical case studies and showcase that the uncovered tree structure enables efficient reductions in the computational burden.

\end{abstract}


\section{Introduction}
\label{sec:introduction}
\par Advances in computational power have enabled the development of large-scale systems in safety-critical domains like smart grids and traffic management \cite{venayagamoorthy2011dynamic,bauza2013traffic}. These systems involve numerous interacting agents with complex or uncertain dynamics \cite{venayagamoorthy2011dynamic,julian2019deep}. While connectivity improves coordination, it also increases complexity and risk \cite{zhang2021physical}. This paper addresses the challenge of scalable safety control and verification, beginning with a review of safety control methods, as scalability challenges stem from the underlying control mechanisms themselves. Control Barrier Functions are employed in \cite{ames2016control,taylor2020learning} to enforce that the controlled system remains within a desirable set, which serves as a characterization of safety. Drawing inspiration from Model Predictive Control (MPC), the authors in \cite{engelaar2024stochastic,hewing2020learning} include risk in the optimization framework. Alongside the aforementioned approaches, there exist formal methods \cite{belta2017formal,tabuada2009verification}, which provide a rigorous framework for reasoning about safety across different levels of abstraction \cite{lamport1983good}. In particular, temporal logic enables systematic specification and verification, offering a structured way to verify probabilistic safety. In this paper, we adopt a correct-by-design synthesis approach \cite{haesaert2020robust} to verify safety in a large-scale probabilistic setting.


 \par Correct-by-design control synthesis for stochastic systems with temporal logic specifications often faces a common challenge: the curse of dimensionality \cite{liu2021symbolic}. This refers to the exponential increase in computational cost as system size grows, making synthesis intractable. While several works \cite{mallik2018compositional,lavaei2021compositional,haesaert2020robust,liu2021symbolic,schon2023verifying} address this issue, they primarily focus on abstraction scalability rather than the computational efficiency of the synthesis algorithm.
 \par Work has also been done in \cite{gorodetsky2018high,rozada2024tensor,ong2015value}  for general dynamic programming problems and in \cite{alora2016automated} temporal logic control to mitigate computational complexity by leveraging low-rank tensor approximations. Although computing probabilistic safety guarantees for temporal logic specifications involves value iteration, value functions and the use of tensor approximations in \cite{gorodetsky2018high, rozada2024tensor, ong2015value, alora2016automated} are fundamentally different and not suitable for computing probabilistic safety guarantees. To the best of our knowledge, no existing methods focus on low-rank tensor computations for probabilistic correct-by-design control synthesis problems.  Therefore, we will explore the structure of the corresponding value functions while also enabling the efficient computation of lower bounds. More precisely, we will enable targeted pruning of the tensor and thereby improve computational efficiency to mitigate the curse of dimensionality.

This paper is organized as follows. 
In the next section, we introduce the problem setup. The computation of the satisfaction of temporal logic specifications via value iterations is followed in Sec. \ref{sec:ValIt} with a novel formulation that enables the quantification of the tensor rank of the value function. In Sec. \ref{sec:extree}, a tree structure is introduced to manage this tensor rank and enable pruning as detailed in Sec.~\ref{sec:optval}. This is followed with numeric case studies in Section \ref{sec:case_study}  and the conclusions and future work (c.f., Sec. \ref{sec:conclusions}). \footnote{Due to space constraints, all proofs are removed in this paper, which are available in the extended version of the paper on arXiv.}

 \par 
 

\section{Preliminaries and problem setup}
In this section, we will present preliminaries on Markov decision processes (MDP), LTL-based task specification, and problem setup. 

\subsection{Models: Markov decision processes (MDP)}
\begin{definition}[MDP]   
  An MDP is denoted as $\mathbf M:=(\mathbb{S},\mathbb{A}, \Tr,s_0 )$ with 
\begin{itemize}
 \item 
 a   state space $\spaceS$ with elements $\vs\in \spaceS$;
 \item an action space $\spaceA$ with elements $\vac\in\spaceA$;
\item a stochastic kernel $\Tr :\spaceS \times \spaceA \times \spaceS \rightarrow  [0,1]$, that assigns to each state-action pair $\vs\in\spaceS$ and $\vac\in\spaceA$ a probability distribution $\Tr(\cdot \mid \vs,\vac)$ over $\mathbb{S}$;
\item an initial state $s_0\in\spaceS$.
\end{itemize} \label{def:normal_mdp}
\end{definition}

\noindent The \textit{execution (state trajectory)} of an MDP is 
\begin{equation}
  \execution_{[0,t]}:= \{ \vs[t]  \mid t=0,1,\ldots \}
\end{equation}
initialized with the initial state $\vs[0]\in \spaceS$. In an execution, each consecutive state, $\vs[t+1]\in\spaceS$, is obtained as a realization $\vs_{t+1} \sim \Tr(\cdot \mid \vs,\vac)$ of the stochastic kernel. 
We say that an MDP is finite if both $\spaceS$ and $\spaceA$ are finite.   We denote with $\mathbf{C}_{\boldsymbol{\pi}}$ a  general control strategy that maps each finite history $s[0],a[0],s[1],a[1],\ldots s[t]$ to an action $a[t].$ A more specific type of control strategy is a Markov policy.
\begin{definition}[Markov policy $\boldsymbol{\pi}$] \label{def:markov_policy}
    A Markov policy $\boldsymbol{\pi}$ is a sequence
    \begin{align}
   \boldsymbol{\pi}:= \{ \pi[t] | t=0,1,\ldots \} \label{eq:def_markov_policy}
    \end{align}
     of   maps $\pi[t]: \spaceS \rightarrow \spaceA   $ that assign an action $a\in \spaceA$ to each state $s\in\spaceS$.   The set of Markov policies is denoted as $\Pi$.
\end{definition}
%
We refer to a Markov policy $\boldsymbol{\pi}$ as defined in Definition~\ref{def:markov_policy} as \textit{stationary} if $\pi[t]$ does not depend on the time index $t$, that is for all $t$ it holds that  $\pi[t]:=\pi$  in \eqref{eq:def_markov_policy} with $\pi:\spaceS\rightarrow \mathbb A$. More precisely, we define a \textit{stationary Markov policy} as
$\boldsymbol{\pi}:=\{\pi,\pi,\ldots\}$.
With a slight abuse of notation, $\pi$ is also used in the rest of the paper to denote a stationary Markov policy.  The set of stationary Markov policies is denoted as $\Pi^s$. 

In this paper, we consider finite MDPs that are abstractions of high-dimensional continuous MDPs. In general, an MDP with a continuous state can also be characterized by stochastic difference equations, such as
\begin{equation} 
\quad x = f(x,u)+w. 
\end{equation}
with the state $x \in \mathbb{R}^{d}$, 
the control input $u \in \mathbb{R}^{d}$, 
and the disturbance $w$ modeled as an independently, identically, distributed white noise $ w \in \mathbb{R}^{d}$, with $w$ drawn from distribution $\mathbb P_w$.  An execution of a stochastic difference equation starts from a given initial state $x[0]$, for each $x[t]$ and $u[t]$ the consecutive states $x[t+1]= f(x[t],u[t])+w[t]$ depends on the realization of $w[t]\sim\mathbb P_w$.

\subsection{Specification}
Consider atomic propositions $p_1,p_2, \dots p_N$ that are true or false. The set of atomic propositions and the corresponding alphabet are denoted by $\AP=\left\{p_1,\dots,p_N\right\}$ and $2^{\AP}$, respectively. Each letter $\letter\in  2^{\AP}$ contains the set of atomic propositions that are true. A (possibly infinite) string of letters forms a word $\word := \letter_0,\letter_1,\dots$, with associated suffix $\word_t := \letter_t,\letter_{t+1},\dots$. The state trajectory $\execution = s[0],s[1],\dots$ of a system is mapped to the word $\word:= L(s[0]),L(s[1]),\dots$ using a given labeling function $L:\mathbb{S}\rightarrow 2^{\AP}$ that translates each state to a specific letter $\letter_t = L(s[t])$. Similarly, execution suffixes $\execution_t:=s[t],s[t+1],\ldots$ 
are translated to word suffixes $\word_t:=\letter_t,\letter_{t+1},\ldots$. 
By combining atomic propositions with logical operators, the language of syntactically co-safe Linear Temporal Logics (scLTL) \cite{belta2017formal} can be defined as follows. 
\begin{definition}[scLTL syntax]
	An scLTL formula $\phi$ over a set of atomic propositions $\AP$, with $p\in \AP$ has syntax
\begin{align}\label{eq:scLTLspec}
\phi ::=  p \,|\, \lnot p \,|\, \phi_1 \wedge \phi_2 \,|\, \phi_1 \lor \phi_2 \,|\, \bigcirc \phi \,|\, \phi_1  \until \phi_2.
\end{align} 
\end{definition} \noindent

The semantics of this syntax can be given for the suffixes $\word_t$. An atomic proposition $\word_t \models p$ holds if $p \in \letter_t$, while a negation $\word_t \models \lnot \phi$ holds if $\word_t \not\models \phi$. Furthermore, a conjunction $\word_t \models \phi_1 \wedge \phi_2$ holds if both $\word_t \models \phi_1$ and $\word_t \models \phi_2$ are true, while a disjunction $\word_t \models \phi_1 \lor \phi_2$ holds if either $\word_t \models \phi_1$ or $\word_t \models \phi_2$ is true. Also, a next statement $\word_t \models \bigcirc \phi$ holds if $\word_{t+1} \models \phi$. Finally, an until statement $\word_t \models \phi_1 \until \phi_2$ holds if there exists an $i\in\mathbb{N}$ such that $\word_{t+i} \models \phi_2$ and for all $j \in \mathbb{N}, 0 \leq j < i$, we have $\word_{t+j} \models \phi_1$.
A system satisfies a specification if 
the generated word $\word_0= L(\boldsymbol{s}_0)$ satisfies the specification, i.e.,  
$\word_0 \models \phi$.


 The focus of correct-by-design control synthesis is to design a controller 
$\mathbf{C}_{\boldsymbol{\pi}}$ for model $\mdpM$ and specification $\phi$, such that the controlled system, $\mdpM \times \mathbf{C}_{\boldsymbol{\pi}}$, satisfies $\phi$. For stochastic systems, we are interested in the \textit{satisfaction probability}, denoted as
\begin{align}
\mathbf{P}(\mdpM \times \mathbf{C}_{\boldsymbol{\pi}} \models \phi).
\end{align}

\subsection{Problem setup}

In this paper, we will consider MDPs with decoupled dynamics. That is, we will consider finite abstractions of stochastic difference equations that are in the following form
\begin{equation}
	\begin{aligned}
		\mathbf{S}:\quad \begin{bmatrix}
			x_1^+
			\\
			\vdots
			\\
			x_m^+
		\end{bmatrix} =   \begin{bmatrix}
			f_1(x_1,u_1)
			\\
			\vdots
			\\
			f_m(x_m,u_m)
		\end{bmatrix}+\begin{bmatrix}
			w_1
			\\
			\vdots
			\\
			w_m
		\end{bmatrix},
		\label{eq:mas}
	\end{aligned}
\end{equation}
with the state $x:=\begin{bmatrix}x_1 & \ldots & x_m\end{bmatrix}^\top \in \mathbb{R}^{\sum_{i=1}^m d_i}$, the control input $u:=\begin{bmatrix}u_1 & \ldots & u_m\end{bmatrix}^\top \in \mathbb{R}^{\sum_{i=1}^m d_i}$, and the state disturbance modeled as white noises  $w:=\begin{bmatrix}w_1 & \ldots & w_m\end{bmatrix}^\top \in \mathbb{R}^{\sum_{i=1}^m d_i}$, $w[t] \sim \mathcal{N}(0,\Sigma)$, where $\Sigma$ is a block diagonal positive symmetric matrix.

\begin{remark}
Many systems can be decoupled into the form of \eqref{eq:mas} by design, or can be approximated into such form. For instance, multi-agent systems (MAS) often exhibit this structure due to their distributed nature. Similarly, networked systems, including large-scale communication and power grids, often possess inherent sparsity or modularity that aligns with this requirement. Moreover, even in systems where full decoupling is not immediately evident, local approximations can often be applied to achieve a similar effect. For example, the paper \cite{liu2021symbolic} discussed how complex systems can be locally approximated as decoupled systems, allowing for computational tractability while preserving essential dynamics.  

\end{remark}

\par In this paper, we study $\mathbf{S}$ as described in \eqref{eq:mas}, which can be abstracted as an MDP $\mdpM$ or equivalently as $m$ independent MDPs $\mdpM^{(i)} = (\mathbb{S}^{(i)},\mathbb{A}^{(i)}, \Tr^{(i)},s_0^{(i)} ), i \in \{1,2,\ldots,m\}$:
\begin{align}
&\mathbf{M} = (\mathbb{S},\mathbb{A}, \Tr,s_0 )  \label{eq:high_mdp}
\end{align}
with  $\spaceS=\prod_{i=1}^{m}\spaceS^{(i)}$, $\spaceA=\prod_{i=1}^{m}\spaceA^{(i)}$, $\Tr=\prod_{i=1}^{m}\Tr^{(i)}$, and $s_0=(s_0^1,\ldots,s_0^m)$. 
\par Similarly, we can define a Markov policy $\boldsymbol{\pi}$ as decoupled if it 
 can be equivalently represented as $m$ independent Markov policies $\boldsymbol{\pi}^{(i)}, i\in \{1,2,\ldots,m\}$, each being the Markov policy of $\mdpM^{(i)}$, such that in Definition~\ref{def:markov_policy}, $\pi[t]$ assigns for the current state $s[t]\in\prod_{i=1}^{m}\spaceS^{(i)}$ an input $a[t]\in \prod_{i=1}^{m}\spaceA^{(i)}$. We refer to the set of decoupled Markov policies as $\Pi_{dec}$.   



%
\begin{assumption0}[Independence requirement]
\label{ass:indAP}
The MDP $\mathbf M$ can be represented by  $m$ independent MDPs $\mdpM^{(i)} = (\mathbb{S}^{(i)},\mathbb{A}^{(i)}, \Tr^{(i)},s_0^{(i)} ),i\in\{1,2,\ldots,m\}$ as in Equation \eqref{eq:high_mdp}. 
 The set of atomic propositions is composed of $m$ sets, that is 
$\AP:=\cup_{i=1}^{m} \AP^{(i)},$
with $\AP^{(i)}\cap\AP^{(j)}=\emptyset,\textmd{ if }i \neq j,\forall i,j \in \{1,2,\ldots,m\}$.
Furthermore, there exists labeling maps 
$L^{(i)}:\spaceS^{(i)}\rightarrow 2^{\AP^{(i)}}$ such that states $(s^{(1)},s^{(2)},\ldots,s^{(m})\in \prod_{i=1}^{m}\spaceS^{(i)}$ are mapped to the sets of atomic propositions as
\begin{align}
L(s^{(1)},s^{(2)},\ldots,s^{(m}):= \bigcup_{i=1}^mL^{(i)}(s^{(i)})\in 2^{\AP}. 
\end{align}
\end{assumption0}
Assumption~\ref{ass:indAP} holds for all theoretical results presented in the remainder of the paper.
	
\textbf{Problem: }    
Consider a high-dimensional MDP $\mathbf M:=(\mathbb{S},\mathbb{A}, \Tr,s_0 )$  \eqref{eq:high_mdp} and an scLTL specification $\phi$ for which  Assumption \ref{ass:indAP} holds. 
Design an algorithm for controller synthesis with guaranteed satisfaction probabilities for $\phi$ that mitigates the curse of dimensionality by leveraging the independence in \eqref{eq:high_mdp} and Assumption \ref{ass:indAP}.

\section{Value iteration for temporal logic specifications}\label{sec:ValIt}
In this section, we first recap the 
computation of satisfaction probabilities for scLTL specifications. 
Afterwards, we quantify the rank of the value functions by leveraging reformulated  dynamic programming operators. 
\subsection{scLTL as a probabilistic reach-avoid problem}
As in \cite{haesaert2020robust,abate2008probabilistic}, the satisfaction of a scLTL specification can be rewritten to  \textbf{a reach-avoid problem} \cite{abate2008probabilistic} by leveraging a Deterministic Finite State Automata (DFA).
\par



\begin{definition}[DFA]
\label{def:DFA}
    A deterministic finite automata is defined by the tuple $\DFA=(Q,q_0,\Sigma_\DFA,\tau_{\DFA},q_f)$.
Here, $Q$, $q_0$, and $q_f$ denote the set of states, initial state, and accepting state, respectively. Furthermore, $\Sigma_\DFA=2^{\AP}$ denotes the input alphabet and $\tau_\DFA:Q\times \Sigma_\DFA \rightarrow Q$ is a transition function.
\end{definition}

For a given word $\word= l_0,l_1,_2,\ldots$, a run of a DFA defines a trajectory $
q_0,q_1,\ldots,q_f 
$
that starts with $q_0$ and evolves according to 
$
q_{t+1} = \tau_{\DFA}(q_t,\letter_t). 
$
The word $\word$ is accepted if the corresponding trajectory ends at $q_f$, that is $\exists t$ such that $q_t =q_f$.
For any scLTL specification $\phi$ based on atomic propositions $\AP$, there exists a DFA $\DFA_{\phi}$ with $\Sigma_{\DFA_\phi}=2^\AP$ such that  $\word$ satisfies the specification, that is $\word\models \phi$, if and only if $\word$ is accepted by $\DFA_{\phi}$. In the sequel, we will generally drop the subscript $\phi$. 
%
%
%
    Consider 
    a finite MDP 
    $\mathbf M:=(\mathbb{S},\mathbb{A}, \Tr,s_0 )$, a DFA $\DFA=(Q,q_0,\Sigma_\DFA,\tau_{\DFA},q_f)$, and a labeling function $L:\spaceS \rightarrow \Sigma_\DFA$. The product $\mdpM_{\DFA}$ is a 
    MDP
\begin{align}
\mdpM_\DFA:=(\bar{\spaceS},\bar{\spaceA},\bar{\Tr},\bar{s}_0)
\end{align}
where the initial state is given as $\bar{s}_0:=(s_0,\tau_\DFA(q_0,L(s_0)))$, and  the state set 
 $\bar{\spaceS}= S\times Q$ is defined for elements  $\bar{s}:=(s,q)$, $\bar{\spaceA}=\spaceA$ is the input set, and $\bar{\Tr}(\cdot | \bar{s},a)$ is the stochastic kernel that assigns for any state $\bar{s}=(s,q)$ and control $a$ a probability to the next state $\bar{s}'=(s',q')$ given as  $\bar{\Tr}(\bar{s}'|\bar{s},a)=\Tr(s'|s,a)$   when $q'=\tau_\DFA(q,L(s'))$, and $=0$ otherwise.

Define the accepting state set of the product model $\mdpM_\DFAs$ as $\bar{\mathbb{S}}_f:=S \times \{q_f\}$. The controlled system $\boldsymbol{\pi} \times \mdpM_\DFAs$ satisfying an scLTL specification $\phi$ can be translated to a reachability problem with the target the accepting state set $\bar{\mathbb{S}}_f$.
More precisely, the control synthesis problem can be defined as the design of a policy $\boldsymbol{\pi}$ such that the probability of reaching the accepting state of $\mdpM_\DFAs$ is optimized, that is, 
\begin{align}
\max_{\boldsymbol{\pi}}\mbP_{\boldsymbol{\pi} \times \mdpM_\DFAs}(\exists t \in \mathbb{N}\cup \infty : \bar{s}_t \in \bar{\mathbb{S}}_f).\label{eq:unbounded_prob}
\end{align}
%
%
%
%
%
%
This unbounded reach problem is 
the limit of the probabilistic bounded reachability 
\begin{equation}
\begin{aligned} 
 \mbP_{\boldsymbol{\pi} \times \mdpM_\DFAs}&(\exists t \in \mathbb{N}\cup \infty : \bar{s}_t \in \bar{\mathbb{S}}_f)
 \\
 &:=\lim _{N \rightarrow \infty} \mbP_{\boldsymbol{\pi} \times \mdpM_\DFAs}(\exists t \leq N: \bar{s}_t \in \bar{\mathbb{S}}_f ). 
\end{aligned}\label{eq:unbounded_probability}
\end{equation}
\noindent {\bfseries Dynamic programming mappings.} 
We define value functions $\tensorV_{q,k}^{\boldsymbol{\pi}}:\spaceS  \rightarrow [0,1]$ as the probability that the accepting state set 
$\bar{\spaceS}_f:=\spaceS\times \{q_f\}$ is reached within $k$ time steps starting from the state $(s,q)$ in $\mathbf M_{\DFAs}$. 
Given a Markov policy $\boldsymbol{\pi}:= \{ \pi[t] | t=0,1,\ldots,N-1\}$, 
value functions can iteratively be computed as 
\begin{equation}
\begin{aligned}
&\tensorV_{q,k+1}^{\boldsymbol{\pi}}(s)= \indicator_{\bar{\spaceS}_f}(s,q)
\\
&\quad +\indicator_{\bar{\spaceS} \setminus \bar{\spaceS}_f}(s,q)  
\expectation^{s',q'}[\tensorV_{q',k}^{\boldsymbol{\pi}}(s') | (s,q),a=\pi(s,q))] 
\label{eq:vi_dfa}
\end{aligned}
\end{equation}
%
with $\pi = \pi{[N\!\! -\!\!1\!-\! k]}$, $k\in\{0,\ldots,N-1\}$, and 
  value functions  initialized as
\begin{equation} \label{eq:value_function_initialization_tensor}
\begin{aligned}
&\tensorV_{q_j,0}(s)&=0, &\text{ for } q_j \in Q\setminus \{ q_f\},    \\
&\tensorV_{q_f,k}(s)&=1, &\text{ for } \,k\,\in \{0,1,..,N-1\}.
\\
\end{aligned}
\end{equation}

\noindent As in \cite{haesaert2020robust,abate2008probabilistic}, the subsequent propositions follow. 



\begin{proposition}[scLTL satisfaction] \label{theorem:bellman_iteration_nonoptimal} For a given policy $\boldsymbol{\pi}=\{\pi[k]|k=0,1,\ldots,N-1\}$ with $\pi[k]:\spaceS \times Q \rightarrow \spaceA$, the satisfaction probability within time horizon $N$ for specification $\phi$ of system $\mathbf{M}$ is defined as 
\begin{equation}
\begin{aligned}
\mbP_{\boldsymbol{\pi} \times \mdpM_{\DFAs}}(\exists t \leq N: \bar{s}_t \in \bar{\mathbb{S}}_f):=&\tensorV_{\bar{q}_0,N}^{\boldsymbol{\pi}}(s_0) 
\end{aligned}
\end{equation} with $\bar q_0= \tau_{\mathcal A}(q_0,L(s_0))$ and 
computed iteratively as in \eqref{eq:vi_dfa}, and initialized as in \eqref{eq:value_function_initialization_tensor}.

\end{proposition}

We can similarly define the computation of the optimal satisfaction probability and the optimal Bellman recursions. 
\begin{proposition}
    [Optimal scLTL satisfaction] \label{cor:optimal_value_func}
   For a bounded horizon $N$, the optimal satisfaction probability for specification $\phi$ of system $\mathbf{M}$ is 
%
 computed iteratively as
   \begin{equation}
   \begin{aligned}
   &\tensorV_{q,k+1}^{\ast}(s)= \indicator_{\bar{\spaceS}_f}(s,q)
\\
&\quad +\indicator_{\bar{\spaceS} \setminus \bar{\spaceS}_f}(s,q)  
\expectation^{s',q'}[\tensorV_{q',k}^{\ast}(s') | (s,q),a=\pi^\ast(s,q)] 
   \end{aligned} \label{eq:optimal_satisfaction_probabilities}
   \end{equation}
  based on  the optimal policy $\pi^*(s,q)$, which is computed as
\begin{align}
\pi^*(s,q)\in \argmax_ {\pi:\mathbb S\rightarrow \mathbb A} \expectation^{s',q'}[\tensorV_{q',k}^{\ast}(s') | (s,q),a=\pi(s,q))] .\notag
\end{align}
\end{proposition}

\begin{proposition}
\label{coro:convergence_value_function}
For an infinite horizon, the optimal satisfaction probability for specification $\phi$ of system $\mdpM$ is
\begin{equation}
\begin{aligned}
\max_\mathbf{\boldsymbol{\pi}}\mbP_{\boldsymbol{\pi} \times \mdpM_{\DFAs}}&(\exists t \in N \cup \infty : \bar{s}_t \in \bar{\mathbb{S}}_f ) 
\\
&=\max\{ \indicator_{q_f}(\bar{q}_0),\tensorV_{\bar{q}_0,\infty}^\ast(s_0)\},
\end{aligned}
\end{equation}
where $\bar{q}_0=\tau_\DFA(q_0,L(s_0))$, and $\tensorV_{\bar{q}_0,\infty}^\ast(s_0)$ is the converged value function defined as 
     $   \tensorV_{\bar{q}_0,\infty}^\ast(s_0) := \lim _{N \rightarrow \infty} \tensorV^{\ast}_{\bar{q}_0,N}(s_0),$
with initialization
$\tensorV^{\ast}_{\bar{q}_0,0}(s_0)=0, $ for $\bar{q}_0 \in Q\setminus \{ q_f\}$,    \\
$\tensorV^{\ast}_{\bar{q}_0,0}(s_0)=1,$ for $ \bar{q}_0 \in \{q_f\}.$
Additionally, the optimal policy is stationary
      $  \boldsymbol{\pi}^\ast =(\pi^\ast,\pi^\ast,\ldots).$
\end{proposition}


\subsection{DFA-informed operators} \label{sec:dfa_informed_operator}
\begin{figure}[ht]
    \centering
    \includegraphics[width=0.5\textwidth]{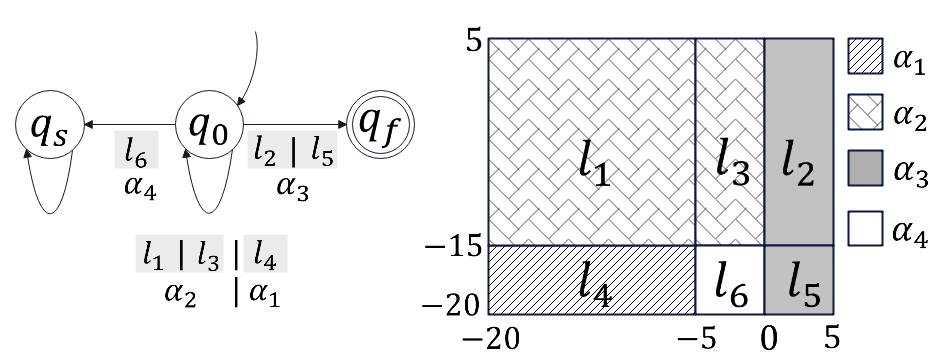}
    \caption{Left: DFA ($q_s$ denotes $q_{\text{sink}}$ for simplicity), right: state space labeled by $\letter$ or $\alpha$.}
    \label{fig:dfa_double_labeled_region}
\end{figure}

In this subsection, we structure the dynamic programming iteration \eqref{eq:vi_dfa} according to the DFA, we call the resulting operators \textit{DFA-informed operators}.  
First, we introduce a running example, based on which we analyze the labelling of the DFA and visualize the DFA-informed operators.  In the sequel, we will leverage the structure in these operators. 
 \begin{example}\label{ex:1}
 	Consider an MDP that is composed from the stochastic difference equations  
 	\begin{equation}\label{eq:ex1} x_{i}^+ = 0.9 x_{i}+0.5 u_{i}+w_{i} \textmd{ for } i=1,2
 	\end{equation}
 	with $w_{i}\sim \mathcal {N}(0,I)$ and with combined state $x=\begin{bmatrix}x_1&x_2\end{bmatrix}^T$.
 	Consider the DFA for the specification $\psi:= (\neg p_2\wedge\neg p_3) \mathsf{U} p_1$ where $p_1, p_2,$ and $p_3$ are atomic propositions for which holds that $p_1 = \operatorname{true} $ iff $x_1\in [0,5]$, $p_2= \operatorname{true} $ iff $x_1\in[-5,0]$, and $p_3 =  \operatorname{true} $ iff $x_2\in [-20,-15]$.
Based on these atomic propositions, the letters combine atomic propositions that are true as $
l_1 = \emptyset, 
l_2=\{p_1\}, 
l_3=\{p_2\}, 
l_4=\{p_3\}, 
l_5=\{p_1, p_3\}, 
l_6=\{p_2, p_3\}$. 
As on the left side of Fig.~\ref{fig:dfa_double_labeled_region}, the transitions of the DFA are triggered by these letters (highlighted under arrows). 
 	Alternatively, we can also label the transitions with Boolean formula (non-highlighted parts under arrows) as depicted on the left side of Fig.~\ref{fig:dfa_double_labeled_region}, that is
 		$\subphi_1  =\neg p_2\wedge p_3 \wedge \neg p_1 \, \Leftrightarrow\{ l_4 \}$, 
 		$\subphi_2 =\neg p_3\wedge \neg p_1\,  \Leftrightarrow\{l_1  , l_3\}$,\,
 		$\subphi_3 =p_1  \Leftrightarrow\{ l_2 , l_5\}$, and
 		$\subphi_4 =p_2 \wedge p_3\, \Leftrightarrow\{l_6\}$.

 \end{example}


\noindent{\bfseries Boolean formula labelled transitions of DFA.}
	For each pair $(q,q^+)\in Q\times Q$, we define a set of Boolean subformulae $\alpha_1, \alpha_2, \ldots, \alpha_{n_\alpha}$ such that iff $l\vDash \bigvee_{i=1,\ldots, n_\alpha} \alpha_i$ then $\tau_\DFA(q,l)=q^+$.  We require that the Boolean formulae are such that
	$\alpha_i$ is composed of a conjunction of (possibly negated) atomic propositions, that is $\alpha::= p|  \neg p | \alpha _1\wedge\alpha_2$, and that  for all $ l\in\Sigma_\DFA$ there exists at most one $i\in\{1,\ldots, n_\alpha\}$ such that $l\vDash\alpha _i$. With a slight abuse of notation we extend the transitions to take these subformulae as arguments, that is $\tau_\DFA(q,\alpha_i)=q^+$.  One can view a subformula being a subset of letters, that is $\subphi_i\subset 2^\AP$.
%
%
%
This is also depicted  in Fig.~\ref{fig:dfa_double_labeled_region} for the DFA of Ex.~\ref{ex:1}. 



 For any DFA $\DFA$ and a set of Boolean formulas, we represent the outgoing transitions of $q$ using \begin{equation}{ \Neig_q}:=\{(\subphi_i,q')| q'=\tau_\DFA(q,\subphi_i)\}.\label{eq:neig_q}
\end{equation}


\par
\begin{figure}[ht]
    \centering
    \includegraphics[width=0.22\textwidth]{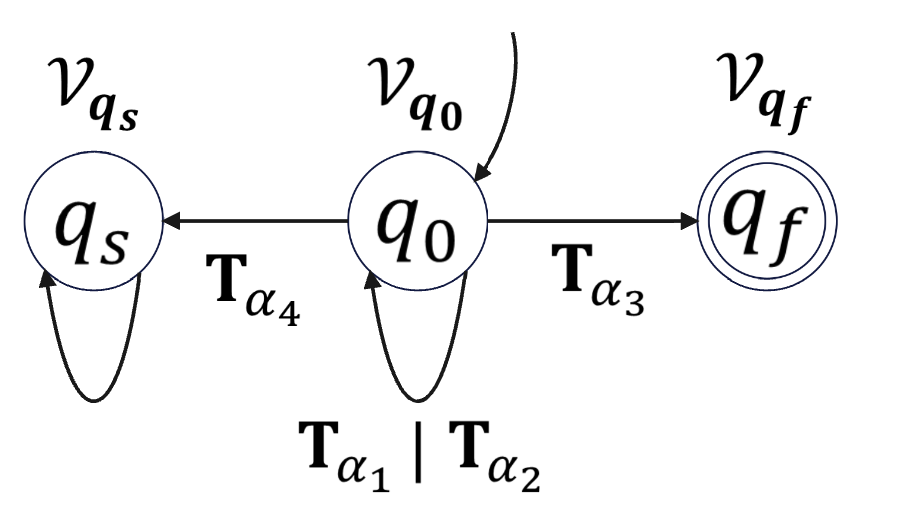}
    \caption{The computation of the value function  based on the operators $\mathbf T_\alpha$ \eqref{eq:OperatorPi} follows the structure of the DFA. The operators that are used to computed $\mathcal V_{q_0}$ are depicted on the edges of the DFA.
    }
    \label{fig:operator}
\end{figure}

 
\noindent{\bfseries DFA-informed operators.}
We denote with $\boldsymbol{\pi}_q(s) = \boldsymbol{\pi}(s,q)$, and define a DFA-informed operator $\op_{\subphi}^{\pi_q}$ for the subformula\footnote{Note that in the remainder we will drop the indices of the subformulae $\alpha_i$.} $\subphi$ and the policy $\pi_q:\spaceS\rightarrow\spaceA$:
\begin{align}\label{eq:OperatorPi}
\op_{\subphi}^{\pi_q}(\tensorV)(s):=\expectation^{s'}[\mathcal{L}_{\subphi}(s')\tensorV(s') | s,a=\pi_q(s)],
\end{align}
with the next state $s'$ evolving based on $\Tr$, and the indicator functions $\mathcal{L}_{\subphi}:\mathbb{S}\rightarrow \{0,1\}$ defined for each subformula $\subphi$ 
as
\begin{equation}
    \mathcal{L}_{\subphi}(s')=\left\{\begin{array}{l}
1, \quad \text{if }L(s') \vDash \subphi\\
0, \quad \text{otherwise.}
\end{array}\right.
\end{equation}
Note that the expected value in \eqref{eq:vi_dfa} is given over the next state $s'$ and the next DFA state $q'$. However, since the DFA has deterministic dynamics in \eqref{eq:OperatorPi} the expected value only includes $s'$.
We then rewrite the value iteration in \eqref{eq:vi_dfa} based on operator $\op_\subphi^{\pi_q}$
as
\begin{align} \label{eq:value_iteration_operator}
\tensorV_{q,k+1}^{\boldsymbol{\pi}}=\sum_{(\subphi_i,q')\in \Neig_q}\op_{\subphi}^{\pi_q[N-1-k]}(\tensorV^{\boldsymbol{\pi}}_{q',k}), \,
\end{align}
for all  $q\in Q\setminus  \{q_f\}$. In Fig.~\ref{fig:operator}, the computation of the value function $\mathcal V_{q_0}$ for Ex. \ref{ex:1} is given based on the operators $\mathbf T_\alpha$ \eqref{eq:OperatorPi} associated to the outgoing transitions of the DFA in Fig.\ref{fig:dfa_double_labeled_region}.

\begin{lemma}[Equivalence of DFA-informed operator]
\label{cor:bellman_operator}
Given a product model $\mdpM_\DFA$ and a policy $\boldsymbol{\pi}$, the value functions $\tensorV_{q,k+1}^{\boldsymbol{\pi}}$ and the corresponding probabilities of reaching the accepting state within time horizon 
$N$ computed based on \eqref{eq:vi_dfa}(Proposition~\ref{theorem:bellman_iteration_nonoptimal}) are equal to those  computed using DFA-informed operators as in \eqref{eq:value_iteration_operator}.
%
%
%
%
\end{lemma}


\begin{lemma}  \label{theorem:optimal_val}
Given a product model $\mdpM_\DFA$, the optimal probabilities of reaching the accepting state within time horizon $N$ computed based on \eqref{eq:optimal_satisfaction_probabilities} (Proposition~\ref{cor:optimal_value_func}) are equivalent to being computed using the proposed DFA-informed operators:
\begin{align}
\tensorV_{q,k+1}^{\ast}={\max_{\pi_q[N-1-k]}}\sum_{(\subphi,q')\in \Neig_q}\op_{\subphi}^{\pi_q[N-1-k]}(\tensorV^\ast_{q',k}) \notag
\end{align}
The optimal policy  for each mode $q$, i.e., $\boldsymbol{\pi}_q^*=(\pi_q^*[0],\pi^*_q[1],\pi^*_q[2],\ldots,\pi^*_q[N-1])$ can be computed as
\begin{align}
\!\!\pi_q^*[N\!-\!1\!-\!k]\in\! \!{\argmax_ {\pi_q[N-1-k]}} 
&\sum_{(\subphi,q')\in \Neig_q}\!\!\!\op_{\subphi}^{\pi_q[N-1-k]}(\tensorV^\ast_{q',k}).
\label{eq:tl_policy_ite}
\end{align}
\end{lemma}

\subsection{Tensor representation of value function}

For $\spaceS=\prod_{i=1}^{m}\spaceS^{(i)}$, 
the value functions $\tensorV_{q,k}^{\boldsymbol{\pi}}:\spaceS  \rightarrow [0,1]$ , can be represented by an order-$m$ tensor, which is a multi-dimensional array, 
\begin{align}
	 \mathbf V \in\mathbb{R}^{n_1\times n_2\times \ldots \times n_m}\label{eq:memo}
	 \end{align}
 with elements $\mathbf V_{i...j}\in[0,1]$. For ease of reasoning we will use  the function $\tensorV_{q,k}^{\boldsymbol{\pi}} $ and its tensor representation $\mathbf V $ interchangeably. 
\begin{figure}[htp]
\begin{subfigure}[b]{0.3\linewidth}
		\includegraphics[width=.9\columnwidth]{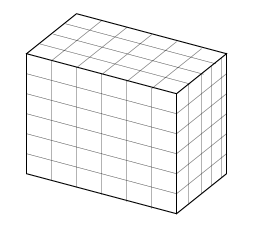}
		\caption{Order-3 tensor with $6\!\times\! 4 \!\times \!6$ elements}
	\end{subfigure}\quad 
\begin{subfigure}[b]{0.6\linewidth}
		\includegraphics[width=\columnwidth]{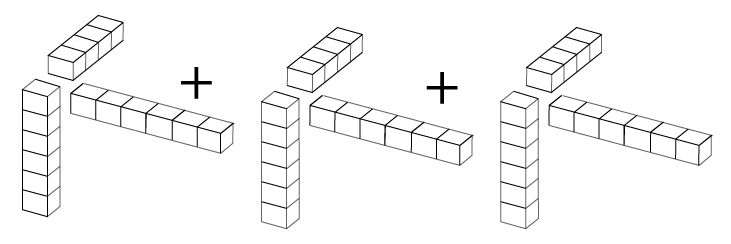}
				\caption{Illustration of Canonical Polyadic Decomposition of Order-3 tensor}
	\end{subfigure}
	\caption{Illustration of order-3 tensor with its Canonical Polyadic Decomposition }\label{fig:CPD}
\end{figure}
Based on Eq. \eqref{eq:memo} the memory required for the value function $\tensorV_{q,k}^{\boldsymbol{\pi}}$ in \eqref{eq:vi_dfa} is  $O(\prod_{i=1}^m n_i)$. However, some tensors can be represented more efficiently by 
their Canonical Polyadic Decomposition.
\begin{definition}[CPD representation] \label{def:CPD_rep}\mbox{ }  Let $\mathbf V\in\mathbb{R}^{n_1\times n_2\times \ldots \times n_m}$ be an order-$m$ tensor. If there exists a Canonical Polyadic Decomposition representation of $\mathbf V$ \cite{hitchcock1927expression}, i.e.,
	\begin{align}
		\mathbf V = \sum_{r=1}^{R} V_{\bullet r}^{(1)} \outerproduct V_{\bullet r}^{(2)} \outerproduct \ldots \outerproduct V_{\bullet r}^{(m)}
	\end{align}
	where $\outerproduct$ denotes the outer product, $V^{(k)}$ is the factor matrix of size $n_k\times R$ for each mode $k$, and $V_{\bullet r}^{(k)}$ is the $r$-th column of $V^{(k)}$, then the CPD rank of $\mathbf V$ is $R$ and is denoted as $\operatorname{rank}(\mathbf V)$.
\end{definition}
As such for a CPD representation with rank $R$, the memory required is of the order $O(R\sum n_i)$.
An order-$3$ tensor is visualized in Fig.~\ref{fig:CPD}, where the tensor rank is $1$. 
%

Note that not all tensors admit a CPD representation. Even for those that do, determining the CPD rank is generally a non-trivial problem \cite{smart2016tensor,kuinchtner2021cp}. 
However, for the studied system $\mathbf{M}$ in \eqref{eq:mas} we know that the dynamics are decoupled and it is to be expected that this will enable a low-rank CPD.
	More precisely,   
for a given policy $\boldsymbol{\pi}\in\Pi_{dec}$, we denote  the controlled transition probability as $\Tr_{\boldsymbol{\pi}}:\spaceS\times\spaceS\rightarrow[0,1]$. The controlled transition probability  inherits  the decoupled nature of both   the transition probabilities and the policy, that is, $\Tr_{\boldsymbol{\pi}}(\prod_i ds_i)=\prod_{i=1}^{m}\Tr^{(i)}_{\boldsymbol{\pi}_i}(ds_i|s_i, a_i).$   

We now show that under Assumption~\ref{ass:indAP}, the value function $\tensorV$ as defined in \eqref{eq:value_iteration_operator}  admits a CPD form, that can be derived directly from the  DFA-informed operators \eqref{eq:value_iteration_operator}. 
\new{

}

\begin{theorem}[Tensor rank of value function] \label{theorem:tensor_rank_val}
 If   Ass.~\ref{ass:indAP} is satisfied then $$\operatorname{rank}(\tensorV_{q,0}^{\boldsymbol{\pi}} )=1 \quad \forall q\in Q.$$ Additionally, for any policy $\boldsymbol{\pi}\in\Pi_{dec}$
and for any value function $\tensorV_{q,k}^{\boldsymbol{\pi}}$ for which 
\begin{align}
\operatorname{rank}(\tensorV_{q,k}^{\boldsymbol{\pi}} )= R_{q,k},   \mbox{ for } q\in Q,\nonumber
\end{align}
  the value function  $\tensorV_{q,k+1}^{\boldsymbol{\pi}}$ computed with \eqref{eq:value_iteration_operator} is such that 
\begin{align}
\operatorname{rank}(\tensorV_{q,k+1}^{\boldsymbol{\pi}})\leq \sum_{(l,q')\in\Neig_{q}}R_{q',k}. \nonumber
\end{align}
\end{theorem}
Theorem~\ref{theorem:tensor_rank_val} bounds the growth of the CPD rank.  Note that the complexity of the DFA and the number of iterations increase the CPD rank of the value function.

\section{Tree-based value iterations }\label{sec:extree}
   
In this section, we propose a tree-based value iteration that supports DFA-informed dynamic programming for high-dimensional systems. This structure facilitates the computation of CPD components of the value function and admits pruning to reduce computational burden.  

\subsection{Trees, values, and tensors}

We first review relevant concepts on graphs and trees \cite{bapat2010graphs}. 
A directed graph can be denoted by $\mathcal{G}=(\setnodes,\mathcal{E})$, where $\setnodes=\{1,\ldots,|\setnodes|\}$ is the set of vertices and $\mathcal{E}\subset \setnodes \times \setnodes$ is the set of edges such that if $(i,j)\in \mathcal{E}$ then the graph contains a directed edge from vertex $i$ to vertex $j$. Vertices, from which there is a directed edge to vertex $j$, are called \textit{parents} of $j$. We call a vertex the \textit{root} vertex if it has no parent, and from it all vertices in the graph can be reached via a sequence of directed edges. We call a directed graph a \textit{rooted tree} if it consists of one root vertex and all the other vertices have one and only one parent. Additionally, we call vertices of a rooted tree with no outgoing edges \textit{leaves}.


\begin{definition}[Tree-based value function $\mathcal{G}$]
A tree-based value function 
  $\mathcal{G}:=\{\setnodes,\mathcal{E},\valuemapping,\qmapping \}$ has a set of vertices $N$ and a set of (labelled) edges $\mathcal{E}:=\{(n,\subphi,n')\}$ (with $\subphi$  the Boolean subformula as in Sec.~\ref{sec:dfa_informed_operator}), which define a \emph{rooted tree}..  Additionally, the functions $\valuemapping$ and $\qmapping$ are the  
vertex value mapping $\valuemapping:\setnodes\rightarrow \mathbb{R}^{\prod |\spaceS^{(i)}|}$  that maps a vertex $n\in \setnodes$ to a tensor $$\valuemapping(n)  \in \mathbb{R}^{\prod |\spaceS^{(i)}|} .$$ 
 and the  DFA-state mapping $\qmapping:\setnodes\rightarrow Q$ that maps a vertex $n\in\setnodes$ to a DFA state $q\in Q$.
\label{def:expanding_tree}
\end{definition}

\begin{algorithm}
\caption{Tree expansion $\mathcal{T}^{\pi}(\mathcal{G})$}\label{alg:sofie}
\begin{algorithmic}[1]
\Procedure{$\mathcal{T}^{\pi}$}{$\mathcal{G}$}
\Comment{Grow tree}
\For{ $n \in \mathcal{G}.\textit{leaves}$} 
\For{  $\{(q,\subphi,q')\in\tau_\DFA|q'=\mathcal{L}_Q(n)\}$}
\State $\bar{n} \leftarrow$ CreateNewVertex
\State $\mathcal G.\setnodes\leftarrow\setnodes\cup \{\bar n\}$ 
\State $\mathcal G.\mathcal{E} \leftarrow \mathcal{E}\cup \{(n,\subphi,\bar{n})\}$
\State $ \mathcal G.\mathcal{L}_Q(\bar{n}) \leftarrow  q$
\EndFor
\EndFor 
\Comment{Update tree values}

\For{  $(n,\subphi,n')\in\mathcal G.\mathcal{E}$}
\State $\valuemapping(n')=\op_{\subphi}^{\pi}(\valuemapping(n))$ 
\EndFor

\EndProcedure
\end{algorithmic}  
\end{algorithm}

Consider  the initial tree-based value function   $\mathcal{G}_0=\{\setnodes_0,\mathcal{E}_0,{\valuemapping}_0,{\qmapping}_0\}$, with $\setnodes_0=\{1\}$, $\mathcal{E}_0=\emptyset$, ${\valuemapping}_0(1)=\indicator$, and $\qmapping(1)=q_f$.  Based on $\mathcal{G}_0$, we can recover the initial value function $\tensorV_{q,0}(s)$ as defined in \eqref{eq:value_function_initialization_tensor}. More precisely,   $\tensorV_{q_f,0}(s)=1$  is represented in $\mathcal{G}_0$ by the  tensor $v_1$ associated  to vertex $1$ in $N_0$.  
\begin{figure}[ht]
	\centering
	\includegraphics[width=0.4\textwidth]{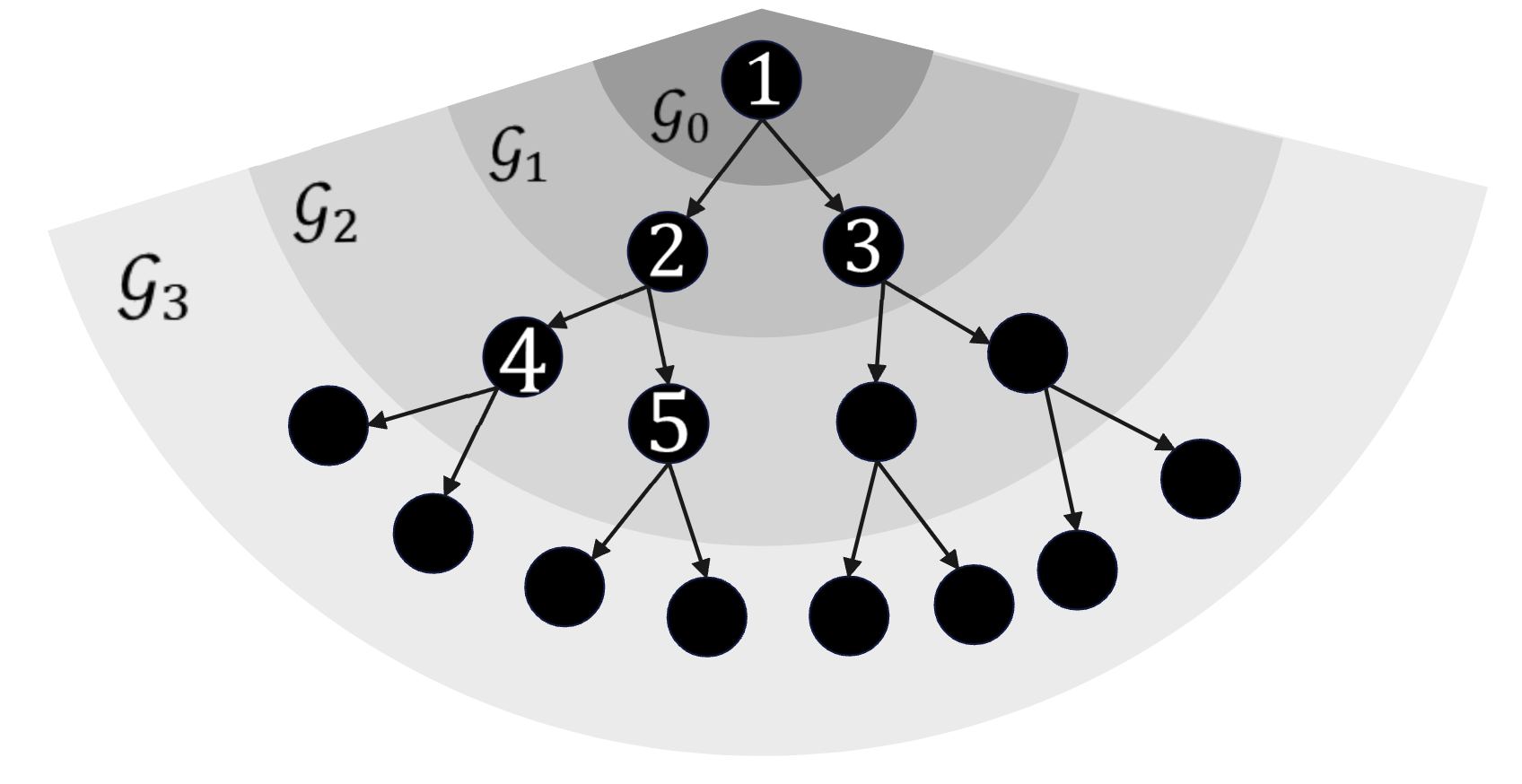}
	\caption{ Tree expanded iteratively based on Alg.~\ref{alg:sofie} as $\mathcal{G}_0$, $\mathcal{G}_1$, $\mathcal{G}_2$, and $\mathcal{G}_3$. 
		Examples: $\mathcal{G}_0$ with $\setnodes_0=\{1\},$ $\mathcal{G}_1$ with $\setnodes_1=\{1,2,3\}$ and $\mathcal{E}_1=\{(1,\subphi,2),(1,\subphi,3)\}$.}
	\label{fig:example_directed_graph}
\end{figure} 
For a given Markov policy
 $   \boldsymbol{\pi} =\{\pi[0],\ldots,\pi[N-1]\},$
the value iteration is equivalent to the expansion of the tree based on operator $\optree^\pi(\mathcal G)$, which -- as described in  Alg.~\ref{alg:sofie} -- includes adding new leaves based on the transitions in the DFA and updating the value mapping $\valuemapping$. More precisely, starting from $\mathcal{G}_0=\{\setnodes_0,\mathcal{E}_0,{\valuemapping}_0,{\qmapping}_0\}$ the three is iteratively grown or expanded
\begin{align}
\mathcal{G}_{k+1}=\optree^{\pi[N-1-k]}(\mathcal{G}_k). \label{eq:tree_expansion_iteration}
\end{align}
%
The tree's growth is also visualized in Fig.~\ref{fig:example_directed_graph}. 


\begin{theorem}
	[Tree mappings to $\tensorV_{q,k}^{\boldsymbol{\pi}}$] \label{theorem:tree_to_value_function} For a horizon N,  a Markov policy 
	$\boldsymbol{\pi}=\{\pi[0],\ldots,\pi[N-1]\}\in\Pi$ 
	and a tree $\mathcal{G}_k=\{\setnodes_k,\mathcal{E}_k,{\valuemapping}_k,{\qmapping}_k\}$ computed with \eqref{eq:tree_expansion_iteration}, we have that 
	\begin{align}
		\tensorV_{q,k}^{\boldsymbol{\pi}}=\sum_{n\in {\qmapping}_k^{-1}(q)}v_k(n), \quad 
	\label{eq:compute_tensorv_based_on_tree}
	\end{align}
	where 
	 ${\qmapping}_k^{-1}(q):Q\rightarrow \setnodes$ gives the vertices set in 
	$\mathcal{G}_{k}$ which is labeled $q$. \eqref{eq:compute_tensorv_based_on_tree} is $\boldsymbol{0}$ when $\{n | \qmapping(n)=q\}=\emptyset$.
\end{theorem}
The above theorem shows how the tree-based value function is connected to the value iterations in Sec.~\ref{sec:ValIt}. Next, we show how the values of the vertices are connected with the CPD representation.
\begin{theorem}[Tensor rank of $v_n$] \label{theorem:tensor_rank_node}
Given a tree $\mathcal{G}$ expanded based on Alg.~\ref{alg:sofie} under Ass. \ref{ass:indAP} and with $\boldsymbol{\pi}\in\Pi_{dec}$, all vertices' values $v(n)$ with $n\in \setnodes$ are of rank $1$, that is,
\begin{equation}
\begin{aligned}
&\exists \mathrm{v}_n^{(1)}\in \mathbb{R}^{|\spaceS^{(1)}|},\mathrm{v}_n^{(2)}\in \mathbb{R}^{|\spaceS^{(2)}|},\ldots \mathrm{v}_n^{(m)}
\in \mathbb{R}^{|\spaceS^{(m)}|}:
\\
&v(n)=  \mathrm{v}_{n}^{(1)}\otimes \mathrm{v}_{n}^{(2)}\otimes \ldots \otimes \mathrm{v}_{n}^{(m)}.\label{eq:node_rank1}
\end{aligned}
\end{equation}
\end{theorem}
Since each subformula can be split over its subsystems as $\subphi=\bigwedge_i \subphi_i$,  we can also define the operator \eqref{eq:OperatorPi} over the subsystem  
\begin{align}
\op_{\subphi}^{	\pi}(v)=\bigotimes_i\op_{\subphi^{(i)}}^{	\pi^{(i)} }(v^{(i)})
\end{align}
if $v=\bigotimes_iv^{(i)}$. 
Together with Theorems \ref{theorem:tree_to_value_function} and \ref{theorem:tensor_rank_node}, the above equation shows that the CPD components of each value function can be recovered as the sum of rank-1 vertices in the tree $\mathcal{G}$.  Via the CPD representation, each rank-1 vertex only requires saving $\sum |\spaceS_i|$ numbers instead of $\prod |\spaceS_i|$.

\subsection{Efficient optimal value iteration}\label{sec:optval}
Although the tree-based value iteration enables the computation of an efficient representation of the value functions, the number of vertices in the tree -- and thus also the rank of the value functions -- can still grow fast. Additionally, the tree is now built for a given policy $\boldsymbol{\pi}\in\Pi_{dec}$.  

\noindent{\textbf{Tree pruning.}}
To improve the computational efficiency of tree-based value iteration, we introduce a tree pruning strategy in which we look for a subtree that includes the root and is connected. 
Given a tree $\mathcal{G}$ with vertices $\setnodes$ and edges $\mathcal{E}$, we define a \emph{subtree} $\hat{\mathcal{G}}$ with vertices $\hat\setnodes\subset\setnodes$ and edges $\hat{\mathcal{E}}\subset\mathcal{E}$ that is a connected graph with same root vertex as $\mathcal{G}$.

\begin{lemma}
	[Lower bound of $\tensorV_{q,k}^{\boldsymbol{\pi}}$ based on pruned tree]\label{lemma:lower_bound_tree}
	Let  $\hat{\mathcal{G}}=(\hat\setnodes, \hat{\mathcal{E}}, \hat{\valuemapping}, \hat{\qmapping})$
	be a subtree of ${\mathcal{G}}=(\setnodes, {\mathcal{E}}, {\valuemapping}, {\qmapping})$ with $\hat\setnodes\subset \setnodes$.   Let $\hat{\tensorV}$ and $\tensorV$ denote the respective value functions computed based on \eqref{eq:compute_tensorv_based_on_tree}. 
	 Then 
	\begin{align}
		\forall s,q: \hat{\tensorV}^{\pi}_{q,k}(s)\leq \tensorV^{\pi}_{q,k}(s).
	\end{align}
	%
%
%
\end{lemma}

We propose a first straightforward approach to obtain a subtree, while doing the value iteration, by pruning the leaves of the tree. More precisely, we choose a threshold value $v_{\text{th}}$ and prune all leaves with whose values fall below it.  This operation $\mathcal{P}(\mathcal G)$ is introduced in \ref{alg:tree_pruning}, and in combination with the tree growing it leads to a more efficient growth as
\begin{align}
\hat{\mathcal{G}}_{k+1} = \mathcal{P}(\mathcal{T}^\pi (\hat{\mathcal{G}_k})).
\end{align}
\begin{algorithm}[htp]
	\caption{Tree pruning $\mathcal{P}(\mathcal{G})$}
	\label{alg:tree_pruning}
	\begin{algorithmic}[1]
		\For{$n \in \mathcal{G}.\textit{leaves}$}
		\For{$v(n) <  v_{\text{th}}$}
		\State $\mathcal{G}.\setnodes \leftarrow \setnodes \setminus \{n\}$
		\State $\mathcal{G}.\mathcal{E}\leftarrow \mathcal{E} \setminus \{\text{parent}(n),\subphi,n\}$
		\EndFor
		\EndFor

	\end{algorithmic}
\end{algorithm}
%
The idea of this strategy is to improve efficiency by only keeping effective vertices in the tree while preserving a lower bound of the satisfaction probabilities.

We want to find policy that gives an as-high-as-possible guarantee on the satisfaction probability. To this end, we need to optimize a decoupled policy $\boldsymbol{\pi}_{dec}\in\Pi_{dec}$ with respect to the value iteration \eqref{eq:tl_policy_ite}.  Note that $\boldsymbol{\pi}_{dec}(s,q)$ is such that the policy is decoupled in $s_1,s_2,\ldots$ for each given $q$. However, the  policy is thus still coupled via the finite modes $q$.  For each mode $q$ and each subsystem $i\in \{1,\ldots,m\}$, we can  develop efficient heuristic or approximate optimization of the policy. In the remainder, we will use the following equation to find an optimized decoupled policy
\begin{equation}
	\begin{aligned}
	\pi_q^{\ast(i)}(s^{(i)},a^{(i)})	\in \arg\max_{\pi_q^{(i)}} \sum_{e\in\mathcal E_{q}} \op_{\subphi^{(i)}}^{	\pi_q^{(i)} }(v^{(i)}_{n}) c_{e}^{(i)}
	\end{aligned}
\end{equation}
where $\mathcal E_{q}:=\{e =(n,\alpha,n')\in \mathcal E| \qmapping(n') = q \}$ and  $$c^{(i)}_{e}:= \prod_{j}^{\{1,\ldots,m\} \setminus \{i\}} \|\op_{\subphi^{(j)}}^{	\pi_q^{(j)} }(v^{(j)}_{n}) \|_1.$$
Alternative approaches could consider the use of sampling based approximations and low-rank approximation of tensors as in \cite{rozada2024tensor,ong2015value}. 
We remark that the above approach only delivers  sub-optimal solutions, which is nevertheless often close to optimal behaviors. This will be discussed in more detail in the following section on case studies. 

%
%

\section{Case studies}\label{sec:case_study}
\par To show the benefits of the tensor-tree-based value iteration approach, we consider several case studies. 
All simulations were run on a laptop computer with a 2.3 GHz 11th Gen Intel Core i7-11800H processor and 16.0 GB of RAM. 
\subsection{Two dimensional case}
\begin{figure}
	\centering
	\includegraphics[width=0.25\textwidth]{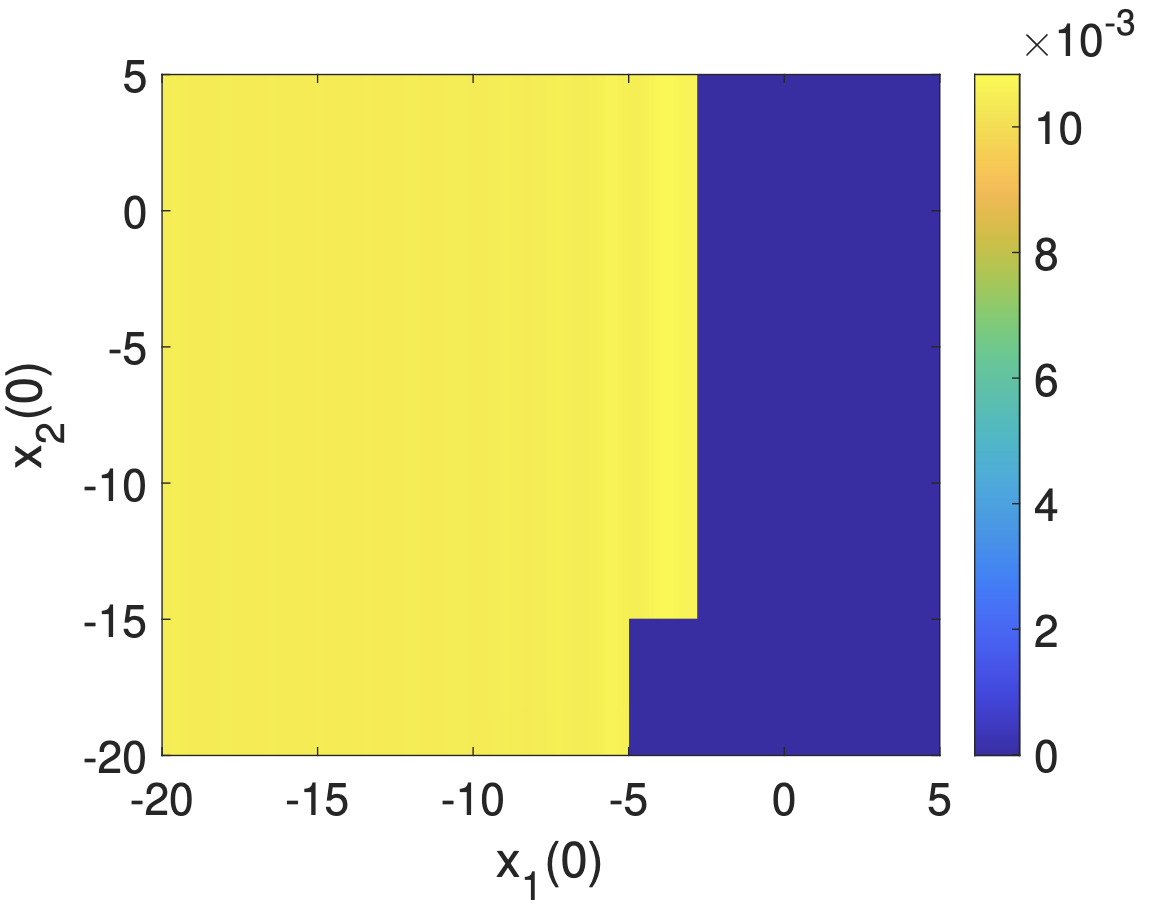}\caption{Satisfaction probability error of rank-1 tree-based value iteration. Error is computed with respect to using exact value iteration.}
	\label{fig:error_rank_1}
\end{figure}

Consider Ex.1 with stochastic difference equations \eqref{eq:ex1} that are gridded as in \cite{haesaert2020robust} to obtain a finite MDP. 
 The specification is the same as in Ex.~\ref{ex:1}: $\psi:= (\neg p_2\wedge\neg p_3) \mathsf{U} p_1$. 
We visualize in Fig.~\ref{fig:error_rank_1} the satisfaction probability error computed as the difference between using exact value iteration and using rank-1 tree-based value iteration. It is shown in Fig.~\ref{fig:error_rank_1} that the maximum approximation error of satisfaction probabilities using rank-1 tree-based value iteration remains within the $10^{-2}$. For each dimension gridded as $1000$ cells, memory usage using rank-$1$ tree-based value iteration is reduced by $90\%$ compared to using exact value iteration, and running time reduced by $90\%$.



\subsection{Four dimensional case}
\begin{figure}[ht]
\centering
     \includegraphics[width=0.45\textwidth]{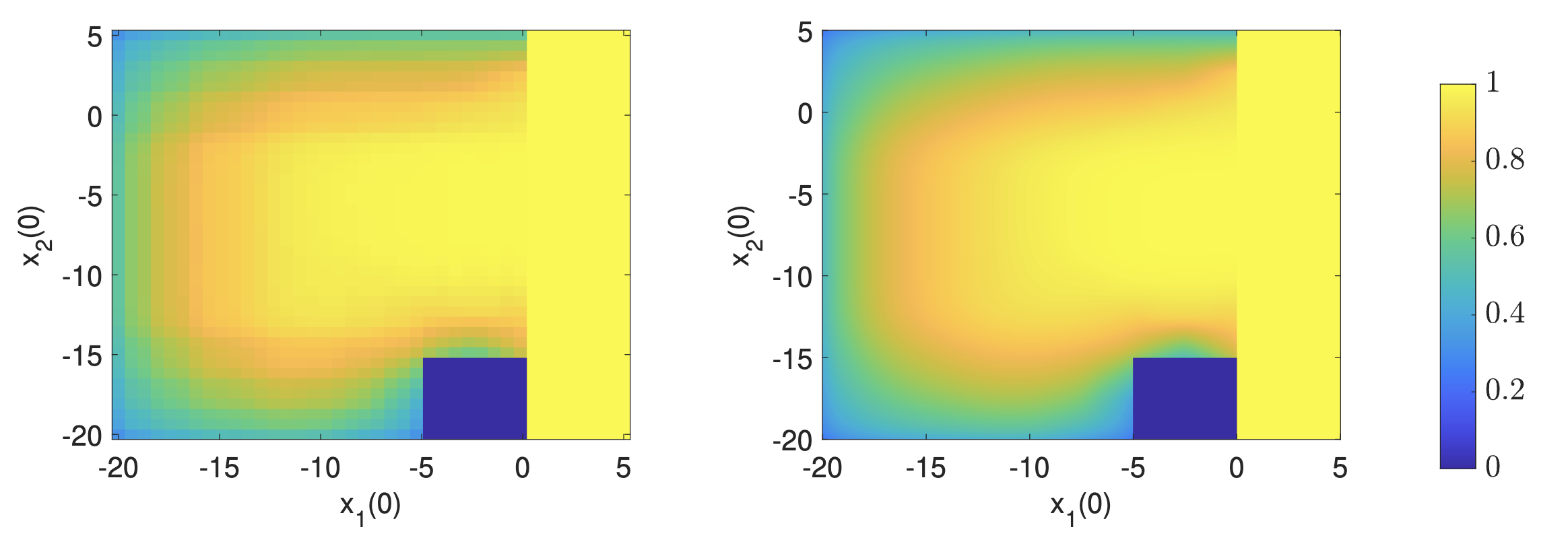}

\caption{Effect of varying number of grid cells visualized as satisfaction probabilities on output space (left: $1.6\times 10^5$ grid cells, right: $10^{12}$ grid cells). } 
\label{fig:4D}

\end{figure}
     
     

\begin{figure}[ht]
    \centering
    \includegraphics[width=0.42\textwidth]{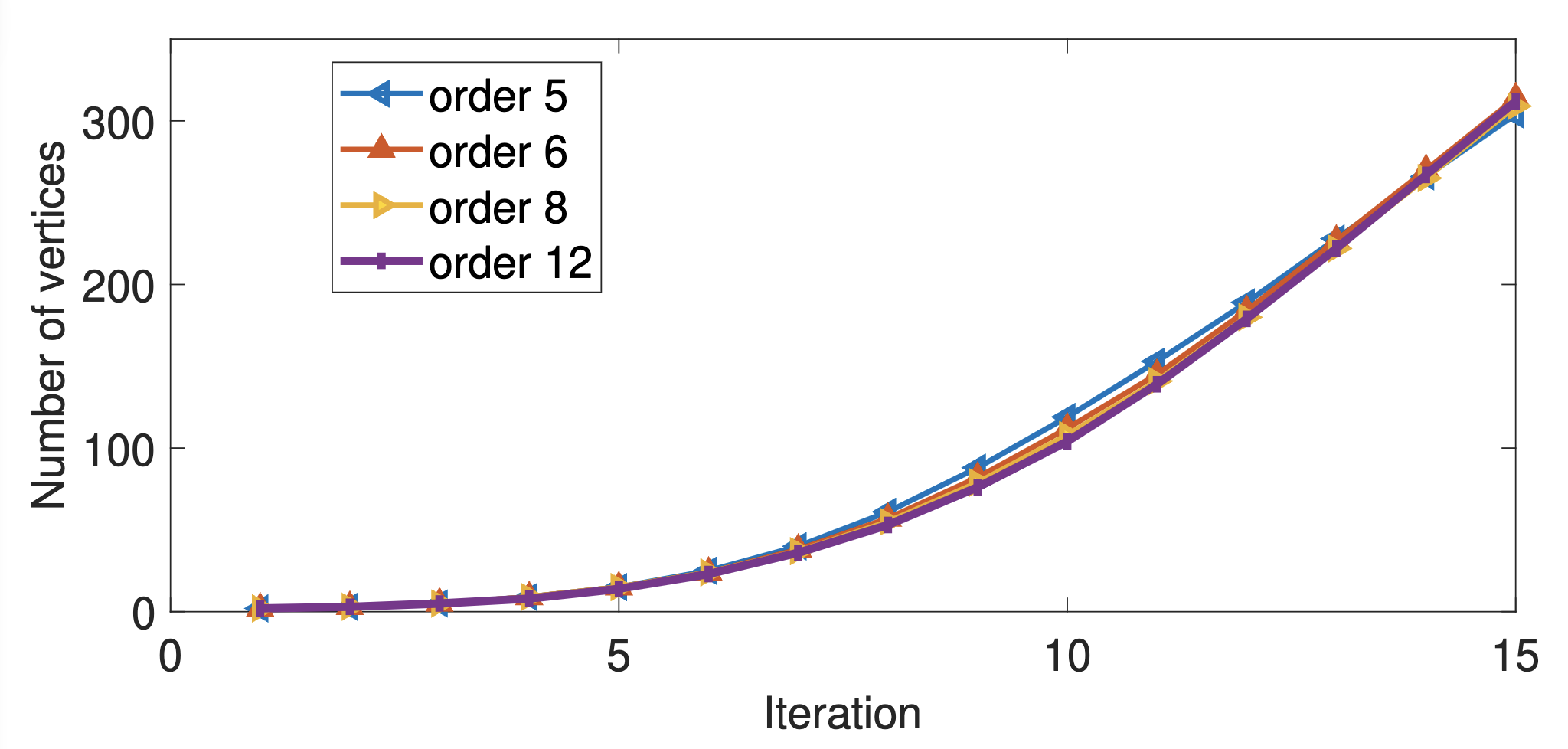}
    \caption{Expanding tree measured by the number of vertices at each iteration step. Each color corresponds to a choice of different order of magnitude in grid cell count.}
    \label{fig:D4_number_nodes}
\end{figure} 
Consider an MDP that is composed from the stochastic difference equations for $i=1,2$ 
\begin{small}
\begin{equation}
\mathbf{S}_i:\quad
\begin{aligned}
\begin{bmatrix}x_1^+\\
x_2^+\end{bmatrix} &=  \begin{bmatrix}1&t_s
\\0&1\end{bmatrix}+\begin{bmatrix}0&0\\1&0
\end{bmatrix}u_1+0.5\begin{bmatrix} w_1\\w_2\end{bmatrix},\nonumber
\\
y_1&=\begin{bmatrix} 1&0\end{bmatrix}\begin{bmatrix}x_1\\x_2 \end{bmatrix}
\end{aligned}
\end{equation}
\end{small}
with $t_s=0.5$, $w_{i}\sim \mathcal {N}(0,1)$, $\spaceX=[-20,5]\times [-5,5] \times [-20,5] \times [-5,5]$, $\mathbb{U}=[-2,2]^2$, $\mathbb{Y}=[-20,5]^2$. The specification is the same as in Ex.~\ref{ex:1}: $\psi:= (\neg p_2\wedge\neg p_3) \mathsf{U} p_1$, where the only difference is that $p_1, p_2,$ and $p_3$ are atomic propositions defined on output space $\mathbb{Y}$ instead of state space $\spaceX$. We validate our tree pruning strategy by computing optimal satisfaction probabilities for a varying number of grid cells, that is $1.6\times 10^5, 5.76\times 10^6, 10^8$, and $10^{12}$. In Fig.~\ref{fig:4D} we visualize such probabilities for state space being discretized as $1.6\times 10^5$ grid cells and as $10^{12}$ grid cells. Running time and memory usage for all 4 cases are summarized in Table~\ref{tab:comp}. With the number of grid cells increased by the order of $8$ (from $10^5$ to $10^{12}$), the memory usage increases by $2.3 \times 10^3$ times (from $0.3482$ MB to $799.8962$ MB). In Fig.~\ref{fig:D4_number_nodes} we showcase that our pruning strategy keeps tree size (in terms of number of vertices) nearly unaffected by the exponential growth of MDP state space (in terms of number of grid cells).



\begin{table}[ht]
\centering
\caption{Performance summary, including running time (in seconds) and memory usage (in MB) of varying system dimensions and grid counts for specification $\psi:= (\neg p_2\wedge\neg p_3) \mathsf{U} p_1$}

    \begin{tabular}{|l|l|l|l|l|}
    \hline
    $d$ & Approach  & Grid count & Time (s) & Memory (MB) 
    \\
    \hline $2$ & Exact VI & $10^6$ & $1.1866$ & $753.3145$ 
    \\
    \hline $2$ & Full rank tree VI  &  $10^6$ & $19.2377$ & $681.3138$ 
    \\
    \hline $2$ & Rank-$1$ tree VI  & $10^6$ & $3.7439$ & $80.2209$ 
    \\
    \hline $4$ & Rank-$1$ tree VI  & $1.6 \times 10^5$ & $3.1862$ & $0.3482$ 
        \\
    \hline $4$ & Rank-$1$ tree VI
     & $5.76\times 10^6$ & $22.069$ & $1.9452$  
        \\
    \hline $4$ & Rank-$1$ tree VI
     & $10^8$ & $28.0122$ & $8.0142$ 
        \\
    \hline $4$ & Rank-$1$ tree VI & $10^{12}$ & $130.5$ & $799.8962$ 
        \\
        \hline
    \end{tabular}
    \label{tab:comp}
\end{table}

\subsection{Scalability benchmark} 

     

\begin{figure}[ht]
    \centering
    \includegraphics[width=0.5\textwidth]{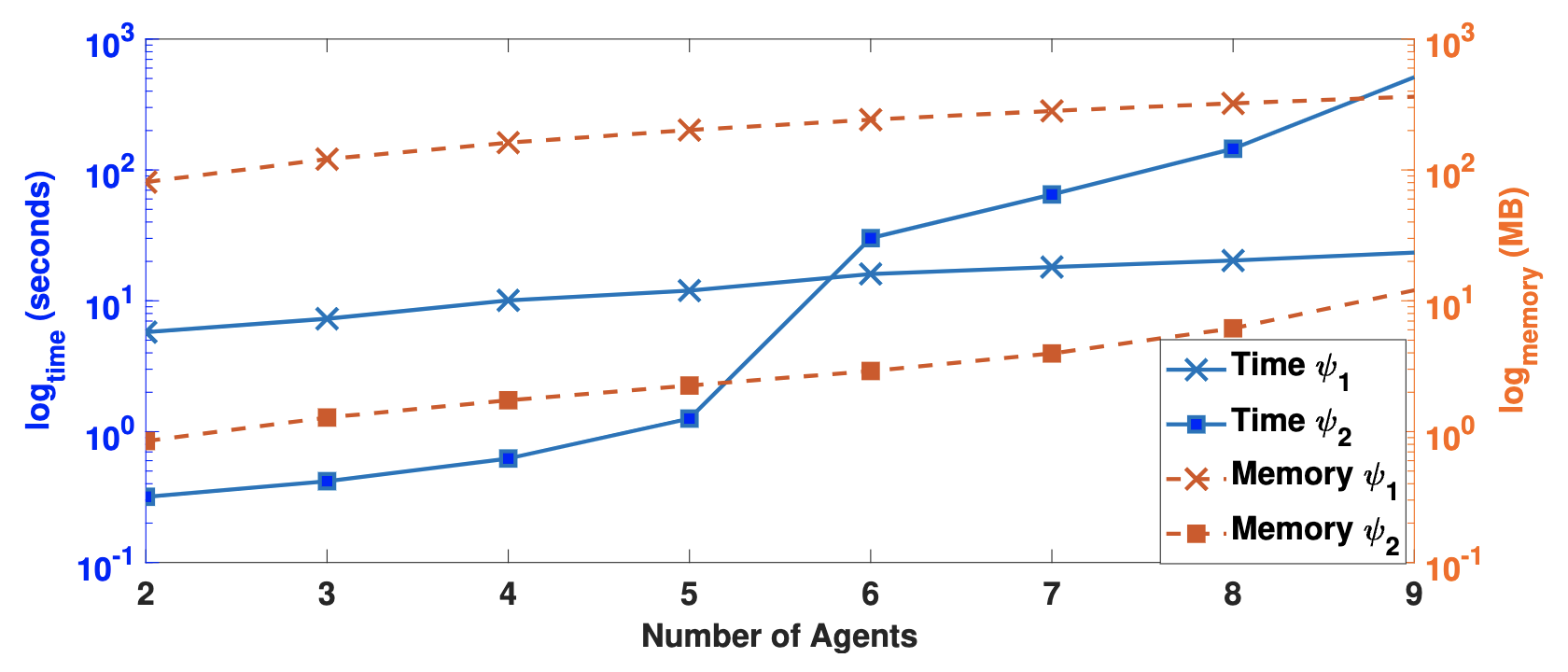}
    \caption{Scalability analysis (on running time in left y-axis as blue and memory usage in right y-axis as orange) for increasing number of agents. \texttt{x} represents time/memory for $\psi_1$, $\blacksquare$ for $\psi_2$. }
    \label{fig:sca}
    
\end{figure}

Consider an MDP that is composed from the stochastic difference equations  
\begin{equation}
    \begin{aligned}
        x_{i}^+ = x_{i}+ u_{i}+w_{i}
    \end{aligned} \label{eq:system_mdp_safety}
\end{equation}
with $w_{i}\sim \mathcal {N}(0,I)$, $\spaceX^{(i)}:[-10,10]$, $\mathbb{U}^{(i)}:[-2,2]$. For scalability analysis purpose, we consider several case studies where the dimension $d$ of the system state varies from $2$ to $9$, that is, from $i=1,2$ to $i=1,\ldots,9$. We consider two two specifications: $\psi_1:=\bigwedge_{t=0}^{5} \bigcirc(\bigwedge_{i=1}^{d}   p_1^{(i)})$ where $p_1^{(i)}$ is atomic proposition for which holds that $p_1^{(i)}=\text{true}$ iff $x_i\in[-5,5]$, and $\psi_2=\bigwedge_{i=1}^{d}p_1^{(i)} \until (\bigvee_{i=1}^{d}p_2^{(i)} \wedge \bigwedge_{i=1}^{d}p_1^{(i)})$, where $p_1^{(i)},p_2^{(i)}$ are atomic propositions for which holds that $p_1^{(i)}=\text{true}$ iff $x_i\in[-5,5]$, and $p_2^{(i)}=\text{true}$ iff $x_i\in[-2,2]$. We consider grid cells up to $10^{27}$ for $\psi_1$, and grid cells up to $10^{18}$ for $\psi_2$. In Fig.~\ref{fig:sca}, the computational time and memory requirements are plotted as functions of the number of agents. Each data point, indicated by the \texttt{x} marker, depicts either the computation time (shown in blue) or memory usage (shown in orange) for the corresponding case study. The results indicate that both computation time and memory usage scale approximately linearly with the number of agents, for invariance specification $\psi_1$, and that memory usage scales approximately linearly for racing specification $\psi_2$.

\section{Conclusion and  future work}\label{sec:conclusions}
This paper presents a  low-rank   value iteration method for correct-by-design control synthesis of stochastic systems under temporal logic specifications. The proposed approach addresses scalability limitations in dynamic programming-based synthesis for large-scale systems via exploiting low-rank tensor structures. A tree-based value iteration approach is proposed that supports DFA-informed dynamic programming for high-dimensional systems. Future work will focus on decomposing complex system dynamics and improving computational efficiency by further reducing the rank of tree-based value iteration.   





\bibliographystyle{IEEEtran}
\bibliography{references}

\newpage
\onecolumn
\appendices
\par 
{\bfseries Proof of Lemma~\ref{cor:bellman_operator}}
\begin{proof}
Initialization in \eqref{eq:value_function_initialization_tensor} can be simplified as
\begin{equation}
\begin{aligned}
\tensorV_{q,0}(s)=\indicator_{q_f}(q).
\end{aligned}
\end{equation}
Denote $\boldsymbol{\pi}(s)=\boldsymbol{\pi}(s,q)$. We unfold \eqref{eq:vi_dfa} by different $q$:
\begin{equation} \label{eq:value_rewrite}
\tensorV_{q,k+1}^{\boldsymbol{\pi}}(s)=\left\{\begin{array}{l}
1, \text{ if } q=q_f\\
\expectation^{s',q'}[\tensorV_{q',k}^{\boldsymbol{\pi}}(s') | (s,q),a=\pi(s,q))], \text{ if } q\neq q_f,
\end{array}\right.
\end{equation}
with $\pi=\pi[N-1-k]$.
Based on the deterministic dynamics of DFA, for the case of when $q \neq q_f$, \eqref{eq:value_rewrite} can be simplified as
\begin{equation}
\begin{aligned}
\tensorV_{q,k+1}^{\boldsymbol{\pi}}(s)
&=\sum_{q'\in Q} \expectation^{s'}[\indicator_{q'}(\tau_{\DFA}(q,L(s')))\tensorV_{q',k}^{\boldsymbol{\pi}}(s') |s,a=\pi_q(s)],
\\
&=\sum_{(\subphi,q')\in \Neig_q}
\expectation^{s'}[\mathcal{L}_{\subphi}(s')\tensorV_{q',k}^{\boldsymbol{\pi}}(s') | (s,q),a=\pi_q(s))],
\end{aligned}
\end{equation}
with $\pi_q=\pi_q[N-1-k]$. We then proved
\begin{equation}
\begin{aligned}
\tensorV_{q,k+1}^{\boldsymbol{\pi}}(s)
&=\sum_{(\subphi,q')\in \Neig_q}\op_{\subphi}^{\pi_q[N-1-k]}(\tensorV^{\boldsymbol{\pi}}_{q',k} (s')) .
\end{aligned}
\end{equation}
\end{proof}
\par
{\bfseries Proof of Theorem~\ref{theorem:tensor_rank_val}}
\begin{proof}
We first prove based on the initialization of the value function defined as in \eqref{eq:value_function_initialization_tensor} that there exists a tensor format of $\tensorV_{q,0}^{\boldsymbol{\pi}}(s_0)$ and the tensor rank is $1$:
\begin{equation}
\begin{aligned}
\tensorV_{q_j,0}&=\mathbf{0}^{(1)} \outerproduct \ldots \outerproduct \mathbf{0}^{(m)},\quad q_j \in Q \setminus \{q_f\}
\\
\tensorV_{q_f,0}&=\indicator^{(1)} \outerproduct \ldots \outerproduct \indicator^{(m)}.
\end{aligned}
\end{equation}
\par The value function is computed iteratively as
\begin{align}
\tensorV_{q,k+1}^{\boldsymbol{\pi}}=\sum_{(\subphi,q')\in \Neig_q}\op_{\subphi}^{\pi_q[N-1-k]}(\tensorV^{\boldsymbol{\pi}}_{q',k})  \label{eq:in_proof_value_iteration}
\end{align}
via the operator $\op_{\subphi}^{\pi_q}$ defined in \eqref{eq:OperatorPi}.
Based on the definition of the operator, the size of $\op_{\subphi}^{\pi_q}(\tensorV)$ is ensured to be the same as the size of $\tensorV$, that is,
\begin{align}
R(\op_{\subphi}^{\pi_q[N-1-k]}(\tensorV^{\boldsymbol{\pi}}_{q',k})) = R(\tensorV^{\boldsymbol{\pi}}_{q',k}),
\end{align}
with $(\subphi,q')\in \Neig_q$. Hence based on \eqref{eq:in_proof_value_iteration}, it is obvious that the tensor rank of $\tensorV_{\bar{q}_0,k+1}^{\pi}$ is computed as the summation of tensor ranks of $\op_{\subphi}^{\pi_{\bar{q}_0}[N-1-k]}(\tensorV^{\pi}_{q',k})$ for all $(\subphi,q')\in \Neig_{\bar{q}_0}$, which is
\begin{align}
\sum_{(\subphi,q')\in\Neig_{q}}R_{q',k}. \nonumber
\end{align}
\end{proof}
\par{\bfseries Proof of Theorem~\ref{theorem:tree_to_value_function}} 
\begin{proof}
We first prove that \eqref{eq:compute_tensorv_based_on_tree} holds for $k=0$. We compute the right-hand side of \eqref{eq:compute_tensorv_based_on_tree} based on $\mathcal{G}_0$ as
\begin{equation}
    \begin{aligned}
        \textmd{for } q=q_f: \sum_{n\in {\qmapping}_0^{-1}(q_f)} v_n&=\valuemapping(1)=1;
        \\
        \textmd{for } q\in Q\setminus \{q_f\}: \sum_{n\in {\qmapping}_0^{-1}(q_f)} v_n&=0.
    \end{aligned}
\end{equation}
Since for the left-hand side of \eqref{eq:compute_tensorv_based_on_tree}, we have
\begin{equation}
    \begin{aligned}
        \textmd{for } q=q_f:\tensorV_{q_f,0}&=1;
        \\
        \textmd{for } q\in Q\setminus \{q_f\}:\tensorV_{q,0}&=0.
    \end{aligned}
\end{equation}
Then we prove \eqref{eq:compute_tensorv_based_on_tree} holds for $k=0$.
\begin{align}
\tensorV_{q,0}=\sum_{n\in {\qmapping}_0^{-1}(q)} v_n.
\end{align}
We then prove that if \eqref{eq:compute_tensorv_based_on_tree} holds for $k$ then it also holds for $k+1$.
\par For $q\not=q_f$, we  rewrite $\tensorV_{q,k+1}^{\boldsymbol{\pi}}$ based on Lemma~\ref{cor:bellman_operator} as
\setlength{\jot}{15pt}
\begin{align}
\tensorV_{q,k+1}^{\boldsymbol{\pi}}&=\sum_{(\subphi,q')\in \Neig_q}\op_{\subphi}^{\pi_{q[N-1-k]}}(\tensorV_{q',k}^{\boldsymbol{\pi}}),  \\
\downarrow&\textmd{\quad \itshape since \eqref{eq:compute_tensorv_based_on_tree} holds for $ k $}\notag
\\
\tensorV_{q,k+1}^{\boldsymbol{\pi}}(s)&=\sum_{(\subphi,q')\in \Neig_q}\op_{\subphi}^{\pi_{q[N-1-k]}}(\sum_{n\in {\qmapping}_k^{-1}(q')}  v_n),  \label{eq:tree_plug_in_sversion_td}
\\
&=\sum_{(\subphi,q')\in \Neig_q}  \sum_{n\in {\qmapping}_k^{-1}(q')} \op_{\subphi}^{\pi_{q[N-1-k]}} (v_n). \label{eq:tree_plug_in_linsum_sversion_td}
\end{align}
The re-writing from \eqref{eq:tree_plug_in_sversion_td} to \eqref{eq:tree_plug_in_linsum_sversion_td} is allowed based on the linearity of the $\sum$ operators. 
Denote the set 
\begin{align} \mathcal N(q) := 
\{(n,\subphi)|   (\subphi,{\qmapping}_k(n))\in \Neig_q \}.
\end{align}
We can then simplify \eqref{eq:tree_plug_in_linsum_sversion_td} to 
\begin{align}
\tensorV_{q,k+1}^{\boldsymbol{\pi}}&=\sum_{(n,\subphi) \in \mathcal N(q)}   \op_{\subphi}^{\pi_{q[N-1-k]}} (v_n).\label{eq:proof:treev3_td}\end{align}
Based on Alg.~\ref{alg:sofie} 
it follows that
\begin{align}
N(q)  = \{(n,\subphi)|(n,\subphi,\Tilde{n})\in\mathcal{E}_{k+1}\wedge q=  {\qmapping}_{k+1}(\tilde n)\}
\end{align}
Based on line 11 of Alg.~\ref{alg:sofie}, we have
\begin{align}
 &\forall (n,\subphi,\Tilde{n})\in\mathcal{E}_{k+1}:\quad v_{\Tilde{n}}=\op_{\subphi}^{\pi_{q[N-1-k]}}(v_n),  
 \\
 &\textmd{ with } v_{\Tilde{n}} \in \mathcal{G}_{k+1} \textmd{ and }v_{{n}} \in \mathcal{G}_{k}
\end{align}
allowing the simplification of \eqref{eq:proof:treev3_td} as
\begin{align}
\tensorV_{q,k+1}^{\boldsymbol{\pi}}(s)
=\sum_{\Tilde{n}\in \setnodes_{q,k+1}} v_{\Tilde{n}}.
\end{align}
where $\setnodes_{q,k+1}:=\{\Tilde{n}\in \setnodes_{k+1} | {\qmapping}_{k+1}(\Tilde{n})=q\}$.
Additionally,
\begin{align}
\tensorV_{q_f,k+1}^{\boldsymbol{\pi}}=1=v_1.
\end{align}
Hence we complete the proof.
\end{proof}
\par{\bfseries Proof of Theorem~\ref{theorem:tensor_rank_node}}
\begin{proof}
We present the proof for the two-dimensional case for simplicity of exposition. The generalization to higher dimensions is immediate from the structure of the argument.
\par Vertices set are initialized as containing only $1$ vertex, that is $\mathcal{G}_0.\setnodes=\{1\}$, value of which is
\begin{equation}
\begin{aligned}
v_1(s_0^{(1)},s_0^{(2)}):=\indicator \outerproduct \indicator
\end{aligned}
\end{equation}
Rank of vertex $1$ is $1$. q-labeling of vertex $1$ is $q_f$.
\par Let vertices $n\in\mathcal{G}_k.\setnodes$ all being rank $1$, with a given policy $\pi$ of rank $1$, the rank of $n' \in \mathcal{G}_{k+1}.\setnodes$ can be computed as follows: (based on line 10 and 11 of Alg.~\ref{alg:sofie} we denote the unique q-labeling of $n'$ is $q'$, of $n$ is $q$ )
\begin{equation}
\begin{aligned}
&v_{n'}(s_0^{(1)},s_0^{(2)})
\\
&:=\indicator_{q}(\tau_{\DFA}(q',L(s_0^{(1)},s_0^{(2)})))
\expectation(v_n(s_0^{(1)+},s_0^{(2)+})),
\\
&=\indicator_{q}(\tau_{\DFA}(q',L^{(1)}(s_0^{(1)}))\outerproduct \indicator_{q}(\tau_{\DFA}(q',L^{(2)}(s_0^{(2)}))
\\
& \quad \sum_{s_0^{(1)+},s_0^{(2)+}} \tensorP^{\pi}_{s_0^{(1)},s_0^{(2)},s_0^{(1)+},s_0^{(2)+}}v_n(s_0^{(1)+},s_0^{(2)+}). \label{eq:node_rank1_unfold_state_trans}
\end{aligned}
\end{equation}
Based on MDPs $\mdpM^{(i)}:=(\spaceS^{(i)},\spaceA^{(i)},\Tr^{(i)})$, each of which is the abstraction of an independent agent, we write state transition probabilities $\tensorP^{\pi}_{s_0^{(i)},s_0^{(i)+}}$ as
\begin{equation}
\begin{aligned}
\tensorP^{\pi}_{s_0^{(i)},s_0^{(i)+}}=\sum_{a^{(i)}} \Tr^{(i)}(s_0^{(i)+} | s_0^{(i)},a^{(i)}) \pi_{s_0^{(i)},a^{(i)}}.
\end{aligned}
\end{equation}
$\tensorP^{\pi}_{s_0^{(1)},s_0^{(2)},s_0^{(1)+},s_0^{(2)+}}$ in \eqref{eq:node_rank1_unfold_state_trans} can be expressed using $\Tr^{(1)}$ and $\Tr^{(2)}$ and written as 
\begin{equation}
\begin{aligned}
&\tensorP^{\pi}_{s_0^{(1)},s_0^{(2)},s_0^{(1)+},s_0^{(2)+}}
\\
&=\sum_{a^{(1)},a^{(2)}}\tensorP_{s_0^{(1)},s_0^{(2)},s_0^{(1)+},s_0^{(2)+}} \pi_{s_0^{(1)},s_0^{(2)},a^{(1)},a^{(2)}},
\\
&=\sum_{a^{(1)},a^{(2)}} \tensorP_{s_0^{(1)},a^{(1)},s_0^{(1)+}} \tensorP_{s_0^{(2)},a^{(2)},s_0^{(2)+}}\pi_{s_0^{(1)},s_0^{(2)},a^{(1)},a^{(2)}}, \label{eq:rank1_trans_rewrite}
\end{aligned}
\end{equation}
where the policy $\pi_{s_0^{(1)},s_0^{(2)},a^{(1)},a^{(2)}}\in \mathbb{R}^{(|\spaceS^{(1)}| \times |\spaceA^{(1)}|)\times(|\spaceS^{(2)}| \times |\spaceA^{(2)}|)}$ is of rank $1$ and can be written as
\begin{equation}
\begin{aligned}
\pi_{s_0^{(1)},s_0^{(2)},a^{(1)},a^{(2)}}=\Pi^{(1)} \outerproduct \Pi^{(2)}, \label{eq:rank1_policy_rewrite}
\end{aligned}
\end{equation}
where vector $\Pi^{(1)}\in\mathbb{R}^{(|\spaceS^{(1)}| \times |\spaceA^{(1)}|)}$, and vector $\Pi^{(2)}\in\mathbb{R}^{(|\spaceS^{(2)}| \times |\spaceA^{(2)}|)}$. 
We re-write $\tensorP^{\pi}_{s_0^{(1)},s_0^{(2)},s_0^{(1)+},s_0^{(2)+}}$ in \eqref{eq:rank1_trans_rewrite} by replacing $\pi_{s_0^{(1)},s_0^{(2)},a^{(1)},a^{(2)}}$ using \eqref{eq:rank1_policy_rewrite}:
\begin{equation}
\begin{aligned}
&\tensorP^{\pi}_{s_0^{(1)},s_0^{(2)},s_0^{(1)+},s_0^{(2)+}}
\\
&=\sum_{a^{(1)},a^{(2)}}\tensorP_{s_0^{(1)},a^{(1)},s_0^{(1)+}} \tensorP_{s_0^{(2)},a^{(2)},s_0^{(2)+}}\pi_{s_0^{(1)},a^{(1)}} \pi_{s_0^{(2)},a^{(2)}}
\\
&=\sum_{a^{(1)}} \tensorP_{s_0^{(1)},a^{(1)},s_0^{(1)+}} \pi_{s_0^{(1)},a^{(1)}}\sum_{a^{(2)}}  \tensorP_{s_0^{(2)},a^{(2)},s_0^{(2)+}} \pi_{s_0^{(2)},a^{(2)}}
\\
&=\tensorP_{s_0^{(1)},s_0^{(1)+}}^{\pi} \tensorP_{s_0^{(2)},s_0^{(2)+}}^{\pi}
\end{aligned}
\end{equation}
We continue to re-write \eqref{eq:node_rank1_unfold_state_trans} by replacing $\tensorP^{\pi}_{s_0^{(1)},s_0^{(2)},s_0^{(1)+},s_0^{(2)+}}$ as:
\begin{equation}
\begin{aligned}
&v_{n'}(s_0^{(1)},s_0^{(2)})
\\
&:=\indicator_{q}(\tau_{\DFA}(q',L^{(1)}(s_0^{(1)}))\outerproduct \indicator_{q}(\tau_{\DFA}(q',L^{(2)}(s_0^{(2)}))
\\
&\quad \times \sum_{s_0^{(1)+},s_0^{(2)+}} [\tensorP_{s_0^{(1)},s_0^{(1)+}}^{\pi} \tensorP_{s_0^{(2)},s_0^{(2)+}}^{\pi} \mathrm{v}_n^{(1)}(s_0^{(1)+}) \outerproduct \mathrm{v}_n^{(2)}(s_0^{(2)+})],
\\
&=\indicator_{q}(\tau_{\DFA}(q',L^{(1)}(s_0^{(1)}))\outerproduct \indicator_{q}(\tau_{\DFA}(q',L^{(2)}(s_0^{(2)}))
\\
&\quad  \times \sum_{s_0^{(1)+},s_0^{(2)+}} \tensorP_{s_0^{(1)},s_0^{(1)+}}^{\pi} \mathrm{v}_n^{(1)}(s_0^{(1)+})  \tensorP_{s_0^{(2)},s_0^{(2)+}}^{\pi} \mathrm{v}_n^{(2)}(s_0^{(2)+}),
\\
&= \indicator_{q}(\tau_{\DFA}(q',L^{(1)}(s_0^{(1)})) \outerproduct \sum_{s_0^{(1)+}} \tensorP_{s_0^{(1)},s_0^{(1)+}}^{\pi} \mathrm{v}_n^{(1)}(s_0^{(1)+}) 
\\
&\quad \times \indicator_{q}(\tau_{\DFA}(q',L^{(2)}(s_0^{(2)})) \outerproduct \sum_{s_0^{(2)+}} \tensorP_{s_0^{(2)},s_0^{(2)+}}^{\pi} \mathrm{v}_n^{(2)}(s_0^{(2)+}),
\\
&=\mathcal{L}_{\subphi^{(1)}}(s_0^{(1)+}) \outerproduct \sum_{s_0^{(1)+}} \tensorP_{s_0^{(1)},s_0^{(1)+}}^{\pi} \mathrm{v}_n^{(1)}(s_0^{(1)+}) 
\\
&\quad \times \mathcal{L}_{\subphi^{(2)}}(s_0^{(2)+}) \outerproduct \sum_{s_0^{(2)+}} \tensorP_{s_0^{(2)},s_0^{(2)+}}^{\pi} \mathrm{v}_n^{(2)}(s_0^{(2)+}),
\\
&=\expectation^{s_0^{(1)+}}[\mathcal{L}_{\subphi^{(1)}}(s_0^{(1)+})\mathrm{v}_n^{(1)}(s_0^{(1)+}) | s_0^{(1)},a^{(1)}=\pi_q^{(1)}(s_0^{(1)})]
\\
&\outerproduct \expectation^{s_0^{(2)+}}[\mathcal{L}_{\subphi^{(2)}}(s_0^{(2)+})\mathrm{v}_n^{(2)}(s_0^{(2)+}) | s_0^{(2)},a^{(2)}=\pi_q^{(2)}(s_0^{(2)})]. \label{eq:rank1_proof_laststep}
\end{aligned}
\end{equation}
\par We propose operators $\op_{\subphi^{(i)}}^{\pi_q^{(i)}} \quad \forall i\in\{1,2,\ldots,m\}$ associated to the operator $\op_{\subphi}^{\pi_q[N-1-k]}$, defined for Boolean formula $\subphi^{(i)}$ and the policy $\pi_q^{(i)}:\spaceS^{(i)}\rightarrow \spaceA^{(i)}$ as 
\begin{equation}
\begin{aligned}
&\op_{\subphi^{(i)}}^{\pi_q^{(i)}}(\mathrm{v}_n^{(i)})(s^{(i)})
\\
&:=\expectation^{s'^{(i)}}[\mathcal{L}_{\subphi^{(i)}}(s'^{(i)}) \mathrm{v}_{n}^{(i)}(s'^{(i)}) | s^{(i)},a^{(i)}=\pi_q^{(i)}(s^{(i)})].
\end{aligned}
\end{equation}
\eqref{eq:rank1_proof_laststep} can be re-written as
\begin{equation}
\begin{aligned}
&v_{n'}(s_0^{(1)},s_0^{(2)})
\\
&:=\op_{\subphi^{(1)}}^{\pi_q^{(1)}[N-1-k]}(\mathrm{v}_n^{(1)}(s_0^{(1)})) \outerproduct \op_{\subphi^{(2)}}^{\pi_q^{(2)}[N-1-k]}(\mathrm{v}_n^{(2)}(s_0^{(2)})),
\end{aligned}
\end{equation}
with $(\subphi^{(i)},q)\in \Neig_{\qmapping(n)}$, q-label of $n'$ assigned as $q$. We proved that $v_{n'}$ are rank-$1$. 
\end{proof}

\end{document}